\documentclass[conference]{IEEEtran}
\IEEEoverridecommandlockouts
\usepackage{cite}
\usepackage{amsmath,amssymb,amsfonts}
\usepackage{algorithmic}
\usepackage{graphicx}
\usepackage{textcomp}
\usepackage{xcolor}
\usepackage{svg}
\usepackage{subfig}
\usepackage{booktabs} % 更好看的表格线
\usepackage{multirow} % 合并单元格
\usepackage{rotating} % For rotated table headers
\usepackage{pifont}
\usepackage{colortbl}
\usepackage{comment}
\usepackage{enumitem}
\usepackage{array}        % For customizing column widths, 
\usepackage{float}
\usepackage{makecell}     % 用于在单元格内换行和设置字体大小
\usepackage{graphicx}     % For the \resizebox command
\usepackage{tabularx}     % 让列宽自动伸缩
\usepackage{subcaption}  
\usepackage{placeins}
\usepackage{pifont}
\usepackage[table]{xcolor}

\usepackage{csquotes}

\definecolor{darkgreen}{rgb}{0.0, 0.5, 0.0}
\definecolor{darkred}{rgb}{0.8, 0.0, 0.0}
\definecolor{darkblue}{rgb}{0.0, 0.0, 0.8}

\def\BibTeX{{\rm B\kern-.05em{\sc i\kern-.025em b}\kern-.08em
    T\kern-.1667em\lower.7ex\hbox{E}\kern-.125emX}}
\begin{document}

\newcommand{\method}{SpecMamba}
\newcommand{\xietong}[1]{{\color{purple}\ [xt: #1]}}
\newcommand{\linfeng}[1]{{\color{blue}\ [lf: #1]}}
\newcommand{\whf}[1]{\textcolor{orange}{[note: #1]}}
\newcommand{\xsq}[1]{\textcolor{green}{[xsq: #1]}}

\title{\method: Accelerating Mamba Inference on FPGA with Speculative Decoding\\
% {\footnotesize \textsuperscript{*}Note: Sub-titles are not captured in Xplore and
% should not be used}
% \thanks{Identify applicable funding agency here. If none, delete this.}
}

% Linfeng Zhong, Songqiang Xu, Huifeng Wen, Tong Xie, Qingyu Guo, Yuan Wang and Meng Li

\author{
\IEEEauthorblockN{Linfeng Zhong$^{1,3*}$, Songqiang Xu$^{1,4*}$, Huifeng Wen$^{1,4}$, Tong Xie$^{1,2}$, Qingyu Guo$^{2}$, Yuan Wang$^{2,5}$, Meng Li$^{1,2,5\dag}$}
\IEEEauthorblockA{$^1$Institute for Artificial Intelligence, Peking University, Beijing, China}
\IEEEauthorblockA{$^2$School of Integrated Circuits, Peking University, Beijing, China}
\IEEEauthorblockA{$^3$School of Electronic and Computer Engineering, Peking University, Shenzhen, China}
\IEEEauthorblockA{$^4$School of Software and Microelectronics, Peking University, Beijing, China}
\IEEEauthorblockA{$^5$Beijing Advanced Innovation Center for Integrated Circuits, Beijing, China}

\thanks{
This work was supported in part by NSFC under Grant 62495102 and Grant 92464104, Grant 62341407, in part by the National Key Research and Development Program under Grant 2024YFB4505004, in part by Beijing Municipal Science and Technology Program under Grant Z241100004224015, and in part by 111 Project under Grant B18001.

$^*$Equal contribution.
$^\dag$Corresponding author.}

}

% \author{\IEEEauthorblockN{1\textsuperscript{st} Given Name Surname}
% \IEEEauthorblockA{\textit{dept. name of organization (of Aff.)} \\
% \textit{name of organization (of Aff.)}\\
% City, Country \\
% email address or ORCID}
% \and
% \IEEEauthorblockN{2\textsuperscript{nd} Given Name Surname}
% \IEEEauthorblockA{\textit{dept. name of organization (of Aff.)} \\
% \textit{name of organization (of Aff.)}\\
% City, Country \\
% email address or ORCID}
% \and
% \IEEEauthorblockN{3\textsuperscript{rd} Given Name Surname}
% \IEEEauthorblockA{\textit{dept. name of organization (of Aff.)} \\
% \textit{name of organization (of Aff.)}\\
% City, Country \\
% email address or ORCID}
% \and
% \IEEEauthorblockN{4\textsuperscript{th} Given Name Surname}
% \IEEEauthorblockA{\textit{dept. name of organization (of Aff.)} \\
% \textit{name of organization (of Aff.)}\\
% City, Country \\
% email address or ORCID}
% \and
% \IEEEauthorblockN{5\textsuperscript{th} Given Name Surname}
% \IEEEauthorblockA{\textit{dept. name of organization (of Aff.)} \\
% \textit{name of organization (of Aff.)}\\
% City, Country \\
% email address or ORCID}
% \and
% \IEEEauthorblockN{6\textsuperscript{th} Given Name Surname}
% \IEEEauthorblockA{\textit{dept. name of organization (of Aff.)} \\
% \textit{name of organization (of Aff.)}\\
% City, Country \\
% email address or ORCID}
% }

\maketitle

\begin{abstract}

% State Space Models (SSMs), particularly Mamba, have emerged as a promising alternative to Transformers for long-sequence modeling, offering superior computational efficiency and scalability. 
% However, deploying Mamba on edge devices faces significant challenges due to its autoregressive decoding bottleneck and the lack of tailored acceleration techniques. 
% While speculative decoding has proven effective for Transformers, its direct application to SSMs is hindered by three key challenges: (1) hidden state backtracking difficulties, (2) incompatibility with tree-based parallel verification, and (3) hardware utilization mismatches between draft and target models.

% To address these challenges, we propose SpecMamba, the first FPGA-based Mamba accelerator with speculative decoding integrating algorithm, schedule, and hardware-level innovations:
% (1) A memory-aware state backtracking strategy that enables hidden state reconstruction while reducing storage requirements by 1$\times$ and communication overhead by 33\% via hybrid on/off-chip storage;
% (2) A one-pass tree inference algorithm enabling full-tree verification in SSMs with 2$\times$ lower latency than baseline methods; 
% and (3) A linear-parallel SSM sequential dataflow architecture co-optimizing memory-bound and compute-bound operations, improving hardware utilization by 40\%.
% Implemented on AMD FPGAs (VHK158/VCK190), SpecMamba achieves 2–3$\times$ speedup and 2–4$\times$ energy efficiency over GPU and prior FPGA solutions (LightMamba) with negligible accuracy loss, demonstrating its potential for edge deployment.

The growing demand for efficient long-sequence modeling on edge devices has propelled widespread adoption of State Space Models (SSMs) like Mamba, due to their superior computational efficiency and scalability. As its autoregressive generation process remains memory-bound,  speculative decoding has been proposed that incorporates draft model generation and target model verification. However, directly applying speculative decoding to SSMs faces three key challenges:
% its direct application faces three key challenges:
(1) hidden state backtracking difficulties, (2) tree-based parallel verification incompatibility, and (3) hardware workload mismatch. To address these challenges, we propose SpecMamba, the first FPGA-based accelerator for Mamba with speculative decoding, which features system, algorithm, and hardware co-design. At the system level, we present a memory-aware hybrid backtracking strategy to coordinate both models. At the algorithm level, we propose first-in-first-out (FIFO)-based tree verification with tiling to minimize memory access. At the hardware level, we customize a dataflow that computes linear layers in parallel and SSM layers in series to enable maximal overlapping. Implemented on AMD FPGA platforms (VHK158 and VCK190), SpecMamba achieves a 2.27$\times$ speedup over GPU baselines and a 2.85$\times$ improvement compared to prior FPGA solutions, while demonstrating 5.41$\times$ and 1.26$\times$ higher energy efficiency, respectively.

\end{abstract}

\section{introduction}
\label{sec:introduction}

State Space Models (SSMs), like Mamba~\cite{Mamba1_gu2023mamba,Mamba2_dao2024transformersareSSMs}, have emerged as a competitive alternative to Transformer-based large language models (LLMs) \cite{touvron2023llamaopenefficientfoundation,OPT_zhang2022opt}. With linear computational complexity, Mamba has exhibited superior demonstrated superior efficiency~\cite{Mamba2_dao2024transformersareSSMs} for long sequence modeling. Consequently, Mamba has undergone extensive development and refinement for various edge application scenarios, such as computer vision~\cite{Vision_liu2024vmamba,Vision_yang2024plainmamba,Vision_zhu2024visionmambaefficientvisual}, multi-modal~\cite{Multi-Modal_wan2025sigma,Multi-Modal_zhao2024cobra,Multi-Modal_dong2024fusionmamba}, medical application~\cite{Medicalwang2024mamba,Medicalyang2024vivim,Medicalguo2024mambamorph,Others3D_medical_image_xing2024segmamba}, and point cloud processing~\cite{PointCloudli20243dmambacomplete,PointCloud_zhang2024point,PointCloud_liang2024pointmamba}.

\begin{figure}[t!]
\centering
% 右对齐
% \hspace*{-30pt}
    % \resizebox{0.60\textwidth}{!}{
\includegraphics[width=\linewidth]{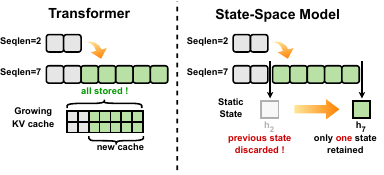}
% }
% \includegraphics[width=6.2cm,height=6cm]{fig/intro.pdf}
\caption{Backtracking during speculative decoding. Transformers store all previous information in KV cache, while SSMs update and discard previous tokens, hindering backtracking.}
% \xietong{what about speculative decoding process of Mamba, to explain hidden states, draft models, backtrack, and so on, or comparing with Transformer speculative decoding process?}
% } 
% \vspace{-10pt}
\label{fig:intro}
\end{figure}

Despite Mamba's impressive efficiency gains, 
its autoregressive generation process remains memory-bound, 
leading to suboptimal computational utilization and thus limited throughput.
% leading to suboptimal computational resource utilization on edge computing platforms such as FPGAs.
% However, research on efficient deployment and acceleration of Mamba models on edge devices remains limited. 
% Existing approaches have made partial progress: MARCA ~\cite{Mamba_MARCA_li2024marca} employs reconfigurable architectures and operator optimization strategies to improve computational and memory access efficiency, while LightMamba~\cite{lightmamba_wei2025lightmamba} enables FPGA deployment through low-precision quantization. 
% Nevertheless, these methods fail to address the fundamental bandwidth bottleneck caused by the sequential nature of Mamba's autoregressive decoding.
% Existing approaches have accelerated Mamba through operator optimization and quantization techniques ~\cite{Mamba_MARCA_li2024marca,lightmamba_wei2025lightmamba}. However, these methods fail to address the fundamental bandwidth bottleneck inherent in Mamba's sequential autoregressive decoding process.
% Inspired by the successful application of speculative decoding in Transformer models~\cite{Speculative_leviathan2023Fast-Inference-via-speculative-decoding,Speculative_chen2023accelerating-large-language-model-decoding-with-speculative-sampling}—where a draft model generates tokens autoregressively while the target model verifies them in parallel—we recognize its potential to increase computational density and alleviate bandwidth constraints without compromising generation quality. 
To mitigate this bandwidth constraint, speculative decoding has been proposed as a promising solution~\cite{Speculative_leviathan2023Fast-Inference-via-speculative-decoding,Speculative_chen2023accelerating-large-language-model-decoding-with-speculative-sampling},
effectively increasing computational density and alleviating bandwidth constraints without compromising generation quality.
This technique employs a two-stage framework: (1) a smaller \textbf{draft model} generates potential token sequences autoregressively, while (2) a larger \textbf{target model} verifies these candidate tokens in parallel. 
Candidate tokens can be further organized into tree structures~\cite{Speculative_miao2024_specinfer} to improve the acceptance rate of the draft tokens and the overall speedup ratio.
% , typically achieving $\geq 2×$ speedup in Transformers.
% This motivates our investigation into adapting speculative decoding to accelerate Mamba inference on edge devices.
% Unlike Transformer-based architectures that utilize a KV cache mechanism, Mamba operates with only a fixed-size hidden state, which dynamically compresses all historical token information.
% While 
% However, Mamba model operates with only a fixed-size hidden state to dynamically compresses all historical token information, it hinders
% leveraging KV cache to select and store previous tokens as in Transformer-based models \cite{Speculative_miao2024_specinfer,Speculative_leviathan2023Fast-Inference-via-speculative-decoding,Speculative_chen2023accelerating-large-language-model-decoding-with-speculative-sampling}.
% Therefore, current researches \cite{SSM_Mamba_Snakes-and-Ladders_wu2024snakes, Mamba-in-llamba_wang2024mamba} primarily focus on  how to (1) enable efficient verification of multiple candidate tokens within a single forward pass, and (2) develop a reliable state recovery mechanism to backtrack to the most recent valid hidden state following token rejection.
However, unlike Transformer-based models that leverage key-value cache (KV cache) to store and select previous tokens, Mamba models dynamically compress historical token information into a fixed-size hidden state, 
%. This mechanism impedes 
impeding backtracking during speculative decoding, as depicted in Fig. \ref{fig:intro}. Therefore, current research \cite{SSM_Mamba_Snakes-and-Ladders_wu2024snakes, Mamba-in-llamba_wang2024mamba} focuses primarily on (1) efficient verification of multiple candidate tokens within a single forward pass and (2) reliable state recovery to backtrack to the last valid hidden state after token rejection.
% how to (1) enable efficient verification of multiple candidate tokens within a single forward pass and (2) develop a reliable state recovery mechanism to backtrack to the most recent valid hidden state following token rejection.

% \subsection{Mamba}
% \label{subsec:mamba}
% \begin{figure}[t!]
% \centering
% \includegraphics[width=\linewidth]{fig/mamba_v2.pdf}
% \caption{The model architecture of Mamba2 and the detailed computation graph of the SSM layer.}
% % \vspace{-15pt}
% \label{fig:mamba}
% \end{figure}

\begin{figure*}[!t]
\centering
\includegraphics[width=\linewidth]{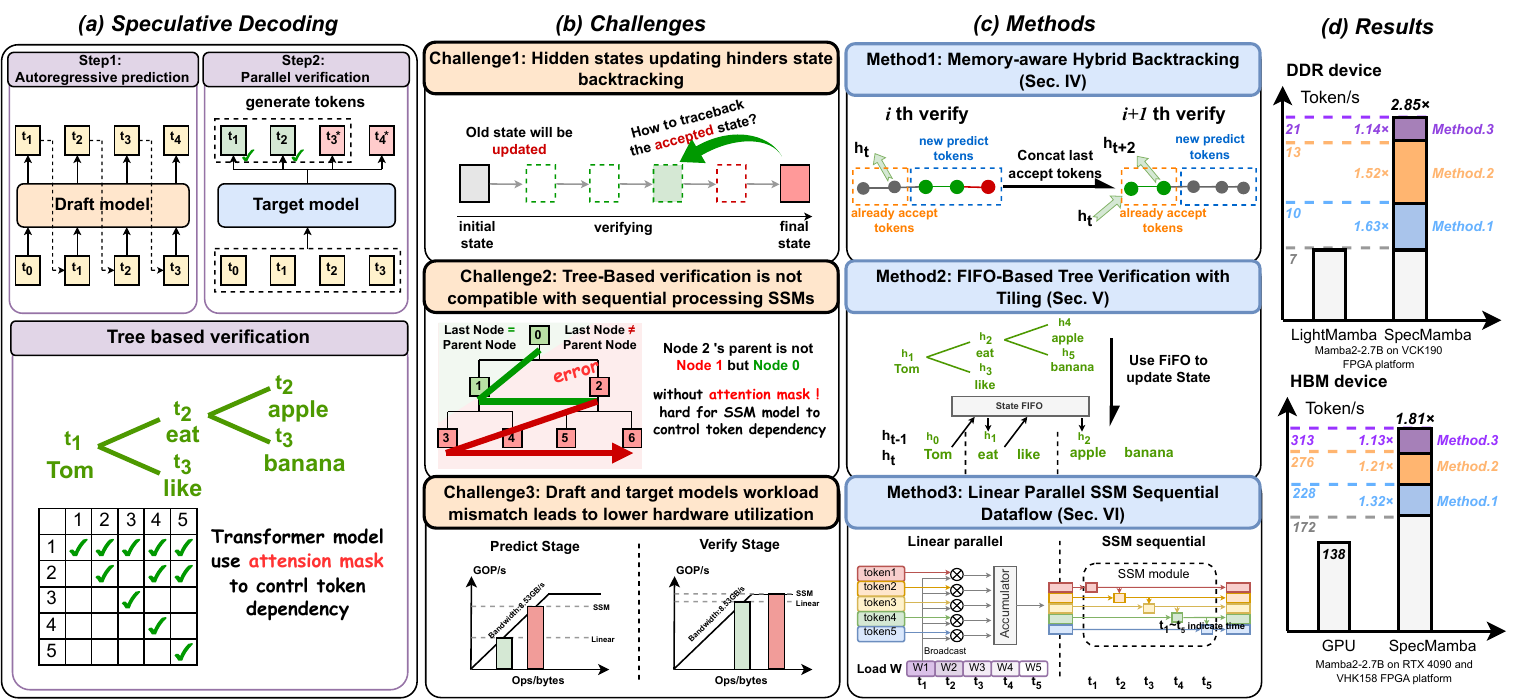}
\caption{Overview of \textit{SpecMamba}. (a)~Dataflow of speculative 
decoding. (b) Challenges of adapting speculative decoding for Mamba inference on FPGA. (c) Methods on \textit{SpecMamba} from Sec.~\ref{sec:method1} to Sec.~\ref{sec:method3}. (d) Results on DDR and HBM devices.} 
\vspace{-10pt}
\label{fig:overall}
\end{figure*}

Despite promising, effectively accelerating Mamba with speculative decoding on edge devices, such as FPGAs, remains an open challenge.
\cite{SSM_Mamba_Snakes-and-Ladders_wu2024snakes,Mamba-in-llamba_wang2024mamba} primarily target GPU platforms and would incur prohibitive overhead if directly deployed on FPGAs. 
\cite{Mamba_MARCA_li2024marca,lightmamba_wei2025lightmamba} have explored Mamba acceleration in edge devices using conventional techniques such as quantization and operator optimization. However, none of them supports speculative decoding. 
Applying speculative decoding to FPGA-accelerated Mamba presents three fundamental challenges, as illustrated in Fig.~\ref{fig:overall} (b):
\textbf{(1) Hidden states updating hinders state backtracking.}
Mamba discards previous hidden states after each update, complicating state recovery when draft tokens are rejected.
\textbf{(2): Tree-based parallel verification is not compatible with sequential processing.} SSMs’ sequential token processing precludes parallel path verification for tree-based drafts via attention masks~\cite{Speculative_miao2024_specinfer} as in Transformers.
\textbf{(3) Draft and target models workload mismatch leads to lower hardware utilization.}
The autoregressive draft model is memory-bound while the parallel verification of target model is compute-intensive. Making both work efficiently on the same hardware without wasting resources is crucial.
% To address the above challenges, we present SpecMamba, a Mamba accelerator with speculative decoding on FPGA, to support fast Mamba computation. 

To address the above challenges, we propose SpecMamba, the first FPGA-based accelerator that supports efficient Mamba inference with speculative decoding. 
As shown in Fig.~\ref{fig:overall} (c), 
At the system level, we present \ding{182}
\textbf{Memory-aware hybrid backtracking}, which combines
off-chip state storage for draft models with on-chip activation
caching for target models.
%, an FPGA-optimized hybrid storage scheme for SSM hidden states, caching intermediate activations on-chip while storing only initial states off-chip. 
% This strategy reduces storage by 3× and communication overhead by 1.68× versus conventional approaches while enabling efficient state reconstruction.
At the algorithm level, we propose \ding{183}
\textbf{First-in-first-out (FIFO)-based tree verification with tiling} that enables tree-based speculative decoding in SSMs by exploiting inter-token dependencies for full-tree verification in a single pass.
% This maintains sequential integrity while achieving 2× latency reduction over baseline methods.
At the hardware level, we employ
\ding{184}
\textbf{Linear Parallel SSM Sequential Dataflow}, 
where we schedule linear layer computation in parallel and SSM computations in series to facilitate overlapping. 
These optimizations achieve significant throughput improvements, as shown in Fig. \ref{fig:overall} (d).
% \whf{Specific results needs to be supplemented.}
% This improves hardware utilization by 40\% over naive implementations.
% we create a linear-parallel dataflow architecture that co-optimizes Mamba's computational patterns with FPGA constraints through memory-bandwidth-aware parallelization and precisely scheduled operation overlapping, yielding 40\% higher hardware utilization.
Our main contributions can be summarized as follows:
% \vspace{-8pt}
\begin{itemize}
    \item We present the first framework that accelerates Mamba inference with speculative decoding on FPGAs. Our framework overcomes the incompatibilities between SSMs' sequential processing and speculative decoding's parallel verification requirements, enabling efficient high-speed inference on edge devices.
    \item We develop a cross-level optimization approach that integrates: (i) a 
    memory-aware hybrid backtracking strategy, (ii) a novel FIFO-based tree verification with tiling algorithm, and (iii) a linear parallel SSM sequential dataflow.
    \item Our implementation on AMD platforms (VHK158 with HBM and VCK190 with DDR) achieves a 2.27× speedup and 5.41× higher energy efficiency compared to GPU baselines. Compared to prior FPGA-based works, SpecMamba demonstrates improvements of 2.85× in execution speed and 1.26× in energy efficiency.

    % . \whf{check}
    % These results validate speculative decoding's effectiveness for Mamba models in resource-constrained scenarios.
\end{itemize}

% We implement SpecMamba on the AMD VHK158 for HBM scenario and AMD VCK190 for DDR scenario. 
% As illustrated in Fig.~\ref{fig:overall}(d), extensive experiments results show that SpecMamba achieves up to 2× and 20 × speedup and 4× and 2× energy enficiency compared with implementation by LightMamba on DDR scenario and GPU on HBM scenario with all the techniques with negligible accuracy loss.

\section{background and related work}
\label{sec:background}

\subsection{Mamba}
\label{subsec:mamba}
\begin{figure}[t!]
\centering
\includegraphics[width=0.95\linewidth]{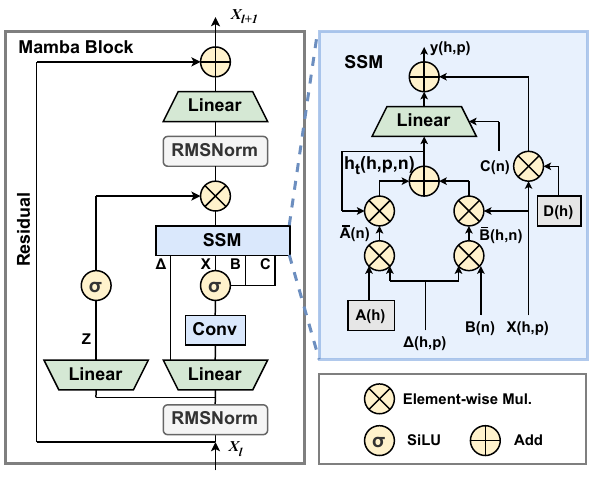}
\caption{The model architecture of Mamba2 and the detailed computation graph of the SSM layer.}
% \vspace{-10pt}
\label{fig:mamba}
\end{figure}

% The Mamba architecture represents a significant advancement in sequence modeling, offering an alternative to attention-based mechanisms by leveraging a structured State Space Model (SSM) formulation. By maintaining a fixed-size hidden state, Mamba decouples computational cost from sequence length, enabling constant-time autoregressive generation. This, in turn, facilitates linear-time inference and significantly reduces memory overhead.

The architecture of Mamba is shown in Fig.~\ref{fig:mamba}. Each Mamba block consists of two linear projection layers  (i.e., an input projection and an output projection), a 1-dimensional convolution layer, a residual connection, an SSM layer for sequence modeling, and two RMSNorm layers.
% During the prefill stage, Mamba summarizes the input prompt 
% The computation of Mamba involves two stages: a prefill stage that summarizes the input prompt, and an autoregressive decode stage that generates output tokens. During decoding, Mamba maintains only a fixed-size hidden state, whereas Transformers store a key-value cache that grows linearly with the sequence length.
% In each Mamba block, the input sequence $X_l$ is first normalized by RMSNorm and projected through a linear layer to produce an intermediate representation denoted as $\Delta$. This representation is then processed by a one-dimensional convolution layer followed by a non-linear activation function, such as SiLU, resulting in two parallel components, B and C. Subsequently, the results is processed by the state space model (SSM) layer. After the SSM operation, the main branch is element-wise multiplied with a gating branch, which is computed via a separate linear projection and activation function (e.g., SiLU). The combined result is further normalized and projected through another linear layer. Finally, a residual connection adds this output back to the original input, yielding the final output of the layer $X_{l+1}$.
In each Mamba block, the input sequence $X_l$ is processed to produce intermediate activations, $\Delta, X, B, C$.
The SSM layer in Eq.~\ref{eq:hidden_state_equation} and Eq.~\ref{eq:y_equation} defines a linear mapping from an input signal $X_t$ to output signal $Y_t$ through a hidden state $h_t$ in each timestep $t$:

% \begin{equation}
%      {h}_t = \bar{A} \odot h_{t-1} + \bar{B} \odot x_t  
% \end{equation}

% \begin{equation}
%      h_t = A h_{t-1} + B X_t \\
% \label{eq:hidden state_equation}
% \end{equation}
% \begin{equation}
%      y_t = C h_{t-1} + D X_t
% \label{eq:y_equation}
% \end{equation}

\vspace{-0.5em} % 减少公式前的间距
\begin{align}
    h_{t} &= \bar{A} \otimes h_{t-1} + \bar{B} \otimes X_{t} \label{eq:hidden_state_equation} \\
    Y_{t} &= C \times h_{t-1} + D \otimes X_{t} \label{eq:y_equation}
\end{align}
% \vspace{-0.5em}% 减少公式后的间距
where $\otimes$ denotes element-wise multiplication.
A typical configuration (e.g., Mamba-2-2.7B with \( h=80\), \(p=64,\) \(n=128 \)) reveals that the hidden state \( h_t \in \mathbb{R}^{h \times p \times n} \) may incur over 1 GB memory overhead during parallel verification, 
thousands of times greater than activations such as \( A \in \mathbb{R}^{h}\), \(B \in \mathbb{R}^{n}\) and \( X \in \mathbb{R}^{h \times p}\).

During SSM processing, the input $X$ undergoes a sequence of transformations. Specifically, $\Delta$ is involved in element-wise multiplications with parameters $A$ and $B$ separately, generating $\bar A$ and $\bar B$. $\bar A$ is then multiplied with the hidden state $h_{t-1}$ and $\bar B$ is element-wise multiplied with input $X$, respectively. The outcomes of these operations are then added to obtain an updated hidden state $h_t$. Finally, $h_t$ is multiplied by $C$, followed by a linear transformation. The result is then added to the element-wise product of the input $X$ and parameter $D$, generating the final output $Y$. 
% The whole computational flow in Mamba block is illustrated in Fig.~\ref{fig:mamba}.

% The SSM layer, the core of the model, captures long-range dependencies through a discretized state-space transformation. The input-dependent step size $\Delta(h, p)$ is processed via Softplus and used to discretize system parameters $A$ and $B$ through element-wise operations. 
% The hidden state evolves recursively as 
% \[
% h_n = A(h, p, n) \odot h_{n-1} + B(h, p, n) \odot x(h, p),
% \]
% while the output is derived as 
% \[
% y(h, p) = C(n) h_n + D(h) \odot x(h, p).
% \]
% % This formulation enables efficient sequential computation while preserving gradient stability, 
% The whole computation flow in Mamba block is illustrated in Fig.~\ref{fig:mamba}.

\subsection{Speculative decoding}
\label{subsec:Speculative decoding}
% Autoregressive decoding in existing LLMs generates one token per iteration based on previously generated tokens, introducing strict sequential dependencies that severely limit parallelism and result in high inference latency. Moreover, the key-value cache used in attention mechanisms incurs substantial memory overhead. To alleviate such inefficiency, speculative decoding~\cite{?} has emerged as a promising and efficient alternative. 

% The key idea behind speculative decoding is that certain decoding steps can be effectively handled by smaller models. 
Speculative decoding leverages smaller language models, referred to as draft models, to generate draft tokens, and adopts the original larger model, named target models, to verify draft tokens in parallel.
When a token fails verification,
% all subsequent tokens are discarded to preserve the correctness of the generated output. This necessitates rolling back to the last verified token and regenerating the remaining sequence. 
the next iteration of draft generation and parallel verification rolls back from the last accepted draft tokens.
% As a result, the overall speedup is limited by the acceptance rate of the draft tokens.
To improve the acceptance rate, recent approaches adopt tree-based verification \cite{OPT-Tree-structure_wang2025opt,li2024eagle1,li2024eagle2,li2025eagle3,cai2024medusa} rather than sequence-based verification \cite{Speculative_leviathan2023Fast-Inference-via-speculative-decoding,Speculative_chen2023accelerating-large-language-model-decoding-with-speculative-sampling}, which construct and verify a tree composed of draft tokens, as illustrated in Fig. \ref{fig:overall} (a). By simultaneously considering more diverse speculation candidates (instead of just one as in sequence-based approaches), tree-based approaches can significantly boost the number of tokens generated in a single forward pass.
As depicted in Fig. \ref{fig:overall} (a), while Transformer-based models use attention masks to restrict tokens 
% (e.g., h5 ("banana")) to attending only their parent nodes (e.g., h1 ("Tom") and h2 ("eats")) 
(e.g., $h_5$ (\enquote{banana})) to attend only their parent nodes (e.g., $h_1$ (\enquote{Tom}) and $h_2$ (\enquote{eats})) in the tree structure, Mamba models lack an analogous mechanism for selective token attention.

To address this, Snakes and Ladders \cite{SSM_Mamba_Snakes-and-Ladders_wu2024snakes} introduced two approaches, the first recovers states by recomputing from augmented input sequences, and the second avoids intensive linear computation by maintaining intermediate activations. \cite{Mamba-in-llamba_wang2024mamba} preserved hidden states for both models, restoring them through selective recomputation during inference. 
However, these methods are designed for GPUs and require extensive tensor transfer, leading to unacceptable communication and memory overhead in bandwidth-limited FPGA platforms.

\begin{table}[t]
    \centering
    \caption{Comparison between different LLM Accelerators.}
    \renewcommand{\arraystretch}{1.0}  
    % \small % 或者 
    \scriptsize
    \label{tab:Comparison of Related LLM Accelerators}
    \resizebox{\linewidth}{!}{
    \begin{tabularx}{\columnwidth}{l>{\centering\arraybackslash}p{1.5cm}>{\centering\arraybackslash}p{1.2cm}>{\centering\arraybackslash}p{1.0cm}>{\centering\arraybackslash}p{1.1cm}}
        \toprule
        & \makecell{\bfseries Mamba \\ \bfseries Compatibility}
         & \makecell{\bfseries Hardware \\ \bfseries Utilization} 
           & \makecell{\bfseries Decode \\ \bfseries Speedup} 
           & \makecell{\bfseries Energy \\ \bfseries Efficiency} 
            \\
        \midrule
        \textbf{\cite{FPGA-accelerator_zeng2024flightllm_daiguohao, background_Pushingup_li2025pushing, Edgellm_huang2025edgellm}} 
        & \textcolor{darkred}{$\times$}
        & \textcolor{darkgreen}{High} & \textcolor{darkblue}{Middle} & \textcolor{darkgreen}{High}  \\
        \textbf{LightMamba\cite{lightmamba_wei2025lightmamba}} 
        & \textcolor{darkgreen}{$\checkmark$}
        & \textcolor{darkred}{Low} & \textcolor{darkblue}{Middle} & \textcolor{darkgreen}{High} \\
        \textbf{MARCA\cite{Mamba_MARCA_li2024marca}} 
        & \textcolor{darkgreen}{$\checkmark$}
        & \textcolor{darkgreen}{High} & \textcolor{darkred}{Low} & \textcolor{darkgreen}{High} \\
        \textbf{Snakes\cite{SSM_Mamba_Snakes-and-Ladders_wu2024snakes}} 
        & \textcolor{darkgreen}{$\checkmark$}
        & \textcolor{darkgreen}{High} & \textcolor{darkgreen}{High} & \textcolor{darkred}{Low}  \\
        \textbf{SpecMamba (Ours)} 
        & \textcolor{darkgreen}{$\checkmark$}
        & \textcolor{darkgreen}{High} & \textcolor{darkgreen}{High} & \textcolor{darkgreen}{High} \\
        \bottomrule
    \end{tabularx}
    }
\end{table}

\begin{figure}[!t]
\centering
\includegraphics[width=0.95\linewidth]{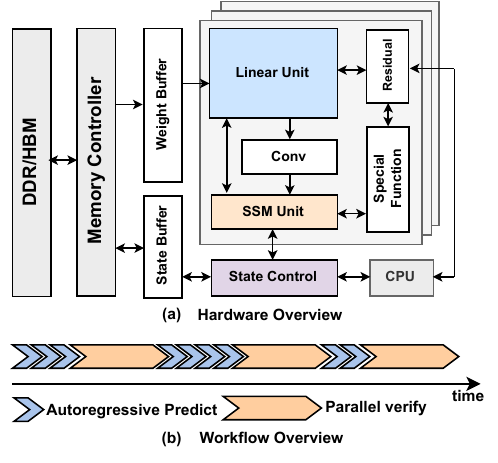}
\caption{SpecMamba overview.} 
\vspace{-10pt}
\label{fig:arch}
\end{figure}

\begin{figure*}[!t]
\centering
\includegraphics[width=\textwidth]{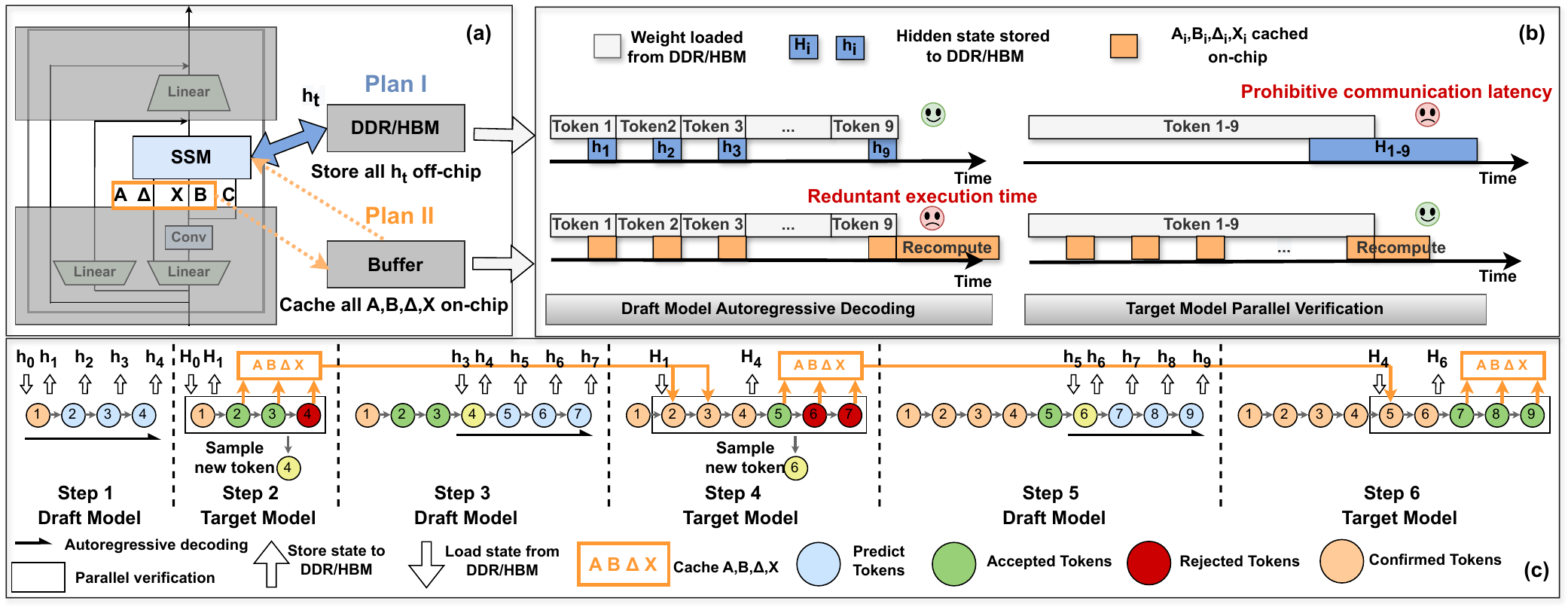} 
\caption{Memory-aware hybrid backtracking. (a) Existing techniques for backtracking. Plan I stores all hidden states off-chip, while Plan II caches lightweight intermediate activations on chip and recomputes hidden states. (b) Tradeoff of Plan I and II for Draft Model and Target Model. For the draft model, Plan II incurs redundant recomputation time. For the target model, Plan I encounters non-overlappable communication overhead. (c) Hybrid backtracking strategy during speculative decoding.} 
\vspace{-10pt}
\label{fig:method1}
\end{figure*}

\subsection{LLM-related Accelerators}
\label{subsec:LLM Accelerators}

As listed in Table \ref{tab:Comparison of Related LLM Accelerators}, 
while traditional techniques such as quantization, sparsification, and operator fusion have demonstrated promising performance for accelerating Transformer-based models \cite{FPGA-accelerator_zeng2024flightllm_daiguohao,background_Pushingup_li2025pushing,Edgellm_huang2025edgellm,MEADOW_moitra2025meadow}, they can not be directly applied to Mamba due to its complex computation graph and distinct activation distribution. 
MARCA \cite{Mamba_MARCA_li2024marca} introduced reconfigurable operators to reconcile the incompatibility between matrix multiplication and element-wise operations, and LightMamba \cite{lightmamba_wei2025lightmamba} focuses on improving decode-phase throughput with rotation-based quantization. However, the bandwidth bottleneck during autoregressive decoding remains unaddressed, leading to severe underutilization of FPGA computation resources.
% integrate rotation-based quantization \cite{} into Mamba and achieve high decode-phase throughput. However, t

% Beyond Transformer-based LLMs, recent research has begun exploring hardware acceleration for emerging state-space models such as Mamba. Marca~\cite{Mamba_MARCA_li2024marca} proposed the first ASIC accelerator for Mamba, introducing reconfigurable operators to reconcile the incompatibility between matrix multiplication and element-wise operations. However, its optimization focus on the prefill stage leaves the bandwidth bottleneck during autoregressive decoding unaddressed. LightMamba~\cite{lightmamba_wei2025lightmamba} subsequently presented the first FPGA-based implementation, achieving high decode-phase throughput through aggressive quantization. Yet due to the inherent memory bandwidth constraints of LLMs, the design utilized only 10\% of the FPGA's compute resources, revealing severe underutilization of processing elements.
Driven by these limitations, we propose SpecMamba, which integrates speculative decoding into an efficient architecture to alleviate memory bandwidth bottlenecks in FPGA-based Mamba inference. By leveraging idle FPGA compute capacity to parallelize draft generation and verification, SpecMamba accelerates the decoding phase and transforms underutilized resources into performance gains. Notably, speculative decoding is orthogonal to existing optimization techniques such as quantization, sparsification, and operator fusion, and thus can be jointly applied to further enhance inference efficiency.

\section{architecture overview}

In this section, we present the overview of \textit{SpecMamba} as in Fig.~\ref{fig:arch} (a), which asynchronously executes draft and target models. This system features a hierarchical memory structure: model parameters are stored in off-chip DDR/HBM while intermediate states and weight tiles are cached within on-chip buffers. 
The design comprises the following components.
\begin{itemize}
    \item Linear Unit: highly parallelized multiplication-and-accumulation (MAC) arrays for matrix operations.
    \item SSM Unit (SSMU): fully unrolled, pipelined element-wise multiplication units (EMUs) for SSM computations.
    \item State Controller: dynamically manages SSM states, issuing memory requests to store and retrieve states.
    \item Memory Controller: schedules data transfers between FPGA and off-chip memory.
    \item Special Function Unit (SFU): accelerates nonlinear functions and feeds results to the SSMU.
    \item Convolution (Conv) Unit: executes 1D convolutions.
    \item Residual Unit: computes residual connections and propagates outputs to subsequent layers or the CPU.
    
\end{itemize}

This unified architecture supports two modes: (1) autoregressive decoding (draft model), dynamically adjusting cycles for variable tree-based computation paths, and (2) parallel verification (target model), validating token trees with single weight loading execution (Fig.~\ref{fig:arch} (b)). This design balances resource efficiency, parallel execution, and adaptive scheduling for speculative decoding.

% \begin{figure}[t]
% \centering
% \includegraphics[width=\linewidth]{fig/arch.pdf}
% \caption{SpecMamba overview.} 
% % \vspace{-10pt}
% \label{fig:arch}
% \end{figure}

% \input{src/motivation}
\section{Memory-aware Hybrid Backtracking
}
\label{sec:method1}

\begin{figure*}[!tb]
\centering
\includegraphics[width=\linewidth]{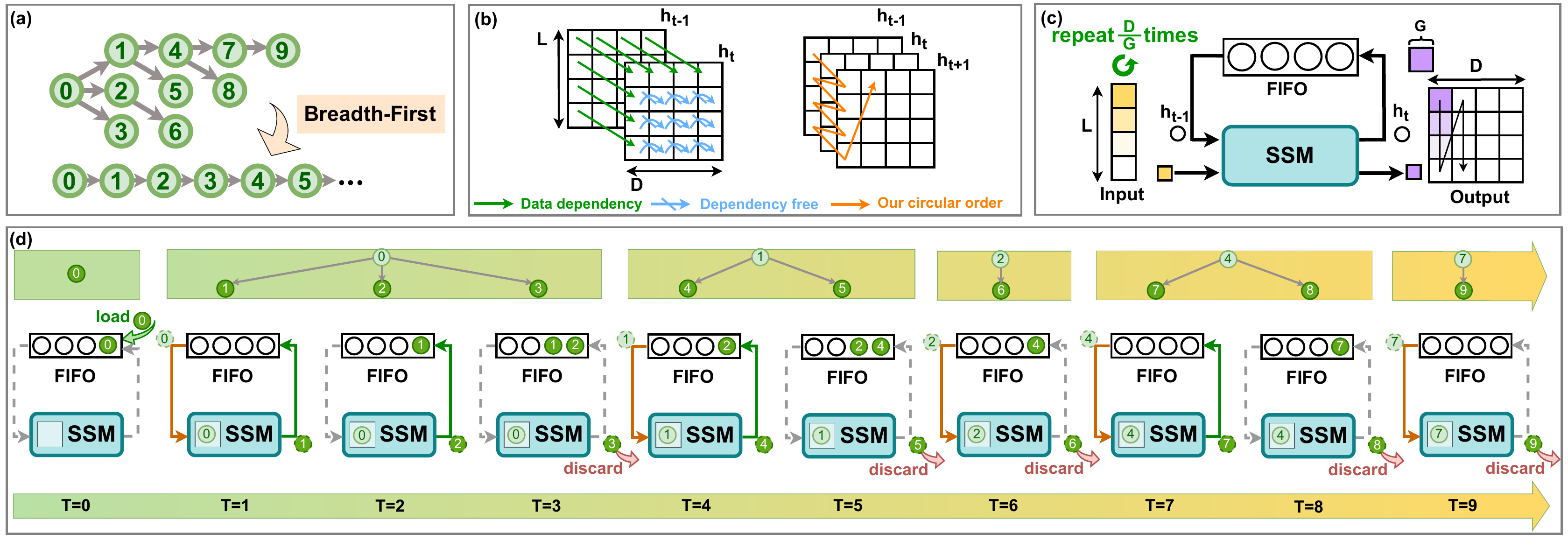}
\caption{FIFO-based tree verification with tiling. (a) Candidate tokens are structured into trees and flattened into sequences via breadth-first traversal. (b) Hidden states have inter-node dependencies, while intra-node tiles are independent, enabling the tiled computation workflow across nodes. (c) Hidden states are managed through an on-chip FIFO that tracks dependencies between tokens. (d) The execution of the tree node hidden state storage and computation is managed by FIFO.} 
\vspace{-10pt}
\label{fig:method2}
\end{figure*}

\subsection{Challenge}
\label{subsec:method1-challenge}
% Unlike Transformers that maintain a growing KV cache, Mamba's hidden state ($h_t$) maintains a constant size regardless of sequence length. While this ensures computational efficiency and memory savings, it introduces a critical challenge for speculative decoding: when draft tokens are rejected, the target model must roll back to the last correct $h_t$ state. However, since Mamba's state updates are strictly causal, previous $h_t$ values are discarded, making traditional backtracking impossible.

% A straightforward solution is to store the $h_t$ for each tokens, which works sufficiently for smaller draft models. However, Take Mamba2-2.7B as an example, a single tokens's $h_t$ already occupies 80{MB}---exceeding the total on-chip BRAM capacity of the VCK190 FPGA. Notably, $h_t$ storage scales linearly with draft sequence length; at a length of 8, it consumes half the memory required by quantized linear layer weights. Even with off-chip storage, this would impose prohibitive communication overhead.

While Mamba achieves memory efficiency by compressing previous tokens into a constant-sized hidden state, this design complicates backtracking in speculative decoding. When draft tokens are rejected, the generation process must roll back to the last accepted token, which is unpredictable before validation. Since Mamba’s casual updates discard prior hidden states, such backtracking becomes difficult.

As shown in Fig. \ref{fig:method1} (a), two main approaches exist to address this. \textbf{Plan I}: storing hidden states of all draft tokens \cite{SSM_Mamba_Snakes-and-Ladders_wu2024snakes}, yet incurring substantial memory overhead. \textbf{Plan II}: caching lightweight activations ($A$, $B$, $\Delta$, $X$) to recompute hidden states \cite{Mamba-in-llamba_wang2024mamba}, which reduces memory overhead but introduces additional computation. Given the distinct workload of draft generation and parallel verification, balancing memory and computation is critical to optimizing FPGA deployments.

\subsection{Observation}
\label{subsec:method1-observation}
% Our motivation stems from a key observation in SSM operations:
% \begin{equation}
%      {h}_t = \bar{A} \odot h_{t-1} + \bar{B} \odot x_t 
%      \label{eq:hidden}
% \end{equation}
% where the dimensions of matrices $\bar{A}$, $\bar{B}$ and $\Delta$ are significantly smaller than $h_t$ itself. 
% By storing only $\bar{A}$ , $\bar{B}$, the memory requirement reduces to merely 670KB per token in the Mamba2-2.7B model. 
% This represents negligible overhead compared to storing the complete $h_t$, enabling us to maintain intermediate activations on-chip while efficiently reconstructing $h_t$ through Eq.~\ref{eq:hidden}.

% By retaining only $\bar{A}$ , $\bar{B}$, and $\Delta$, the storage requirement reduces to 670{KB} for single token. 

% As detailed in Section~\ref{sec:background}, prior work proposed three methods for $h_t$ backtracking \xietong{not mentioned in Sec. II}. However, these implementations target general-purpose GPUs and fail to account for the constraints of edge devices with fixed operators and limited bandwidth. When deployed on FPGAs, they introduce substantial additional computation and communication costs. Table~\ref{tab:comparison} quantitatively compares these approaches against our solution.
% \xietong{describe the shortage of baselines, memory, communication and computation}

Mamba's hidden state generation involves linear projection and SSM computations. Linear layer weights are typically larger than hidden states \cite{Mamba2_dao2024transformersareSSMs}, which requires tiled off-chip loading per execution. SSM operations incur minimal memory access. Fig. \ref{fig:method1} (b) contrasts Plan I (store all states) and Plan II (recompute states) for draft and target models.
For draft model autoregressive decoding with one weight load per token, Plan I enables hidden state loading overlapping with weight transfers, eliminating latency overhead. Plan II introduces recomputation penalties during hidden state regeneration. Conversely, target model parallel verification executes linear layers once but updates hidden states $H_t$ via SSMs sequentially. Here, multiple hidden state storage of Plan I creates prohibitive communication latency, while hidden state recomputation in Plan II optimizes throughput. 
This analysis demonstrates the critical trade-off: Plan I benefits draft models through memory-transfer overlapping, but incurs prohibitive storage overhead for target models. Plan II optimizes target model verification throughput, yet sacrifices draft generation efficiency due to recomputation costs.

% \begin{figure*}[!tb]
% \centering
% \includegraphics[width=\linewidth]{fig/Method2_new.pdf}
% \caption{FIFO-based tree verification with tiling. (a) Candidate tokens are structured into trees and flattened into sequences via breadth-first traversal. (b) Hidden states have inter-node dependencies, while intra-node tiles are independent, enabling the tiled computation workflow across nodes. (c) Hidden states are managed through an on-chip FIFO that tracks dependencies between tokens. (d) The execution of the tree node hidden state storage and computation is managed by FIFO.} 
% % \vspace{-10pt}
% \label{fig:method2}
% \end{figure*}

\subsection{Approach}
\label{subsec:method1-approach}

Based on these insights, we present a memory-aware hybrid hidden state backtracking strategy in Fig. \ref{fig:method1} (c) that combines off-chip state storage for draft models with on-chip activation caching for target models. In step 1, the draft model generates tokens $t_1\sim t_4$, storing hidden states  $h_1\sim h_4$ off-chip. In step 2, the target model stores hidden state $H_1$ off-chip while caching activations $A$, $B$, $\Delta$, $X$ on-chip, then rejects $t_4$ and resamples. In step 3, the draft model resumes from $h_3$ to generate tokens $t_4\sim t_7$, storing $h_4\sim h_7$. In step 4, the target model recomputes $H_2\sim H_4$ using cached activations, with $H_4$ stored off-chip to enable incremental recomputation in step 6. This execution pattern persists throughout subsequent steps, optimizing backtracking latency through strategic balancing of computational and memory costs.

\section{FIFO-based Tree Verification with Tiling}
\label{sec:method2}

% \xietong{ FIFO and Tiled Execution}
% \begin{figure*}[t!]
% \centering
% \includegraphics[width=\linewidth]{fig/method2.pdf}
% \caption{method2} 
% \vspace{-10pt}
% \label{fig:method2}
% \end{figure*}

% \begin{enumerate}[label=(\arabic*), ref=(\arabic*)]
% \item Broken causality: Parallel validation of tree branches requires non-monotonic state dependencies, complicating hardware-friendly execution.
% \item State persistence: Intermediate $h_t$ cannot be discarded immediately, as later leaf nodes may depend on them, necessitating costly scheduling and memory management.
% \end{enumerate}

\subsection{Challenge}
% While the first method enables efficient $h_t$ backtracking in SSMs, 
% further improving the draft acceptance rate requires tree-based speculative decoding. 
% Unlike Transformers, where attention masks natively restrict each token to attend only to its parent in the tree. 
% SSMs lack such a mechanism. 

% % While tree-based speculative decoding can effectively improve the draft acceptance rate, SSMs lack a mechanism like attention masks in Transformers where tokens are restrict
% As discussed in Sec. \ref{}, SSMs lack a mechanism to restrict tokens to attending only their parent nodes.
% This poses a critical challenge: validating tree-structured drafts in a single forward pass without violating SSMs’ sequential dependencies.
% Tree-based drafts demand flattening the tree into a sequence as illustrated in Fig.~\ref{fig:method2}(a), which disrupts the causal structure: sibling tokens no longer share the same parent state, forcing redundant recomputation or state storage. This introduces broken causality: Parallel validation of tree branches requires non-monotonic state dependencies. Therefore, intermediate hidden states of tokens have to be maintained in either BRAM or HBM/DDR, as later leaf nodes may depend on them.
% This can lead to unacceptable on-chip memory overhead or significant  communication overhead, respectively.
A critical limitation of SSMs is their lack of mechanisms to enforce parent-only token attention. This creates a fundamental conflict in tree-structured draft verification: while SSMs inherently require sequential state dependencies, verifying flattened tree drafts (Fig.~\ref{fig:method2} (a)) disrupts causal structure. 
% Sibling tokens lose shared parent-state inheritance, necessitating redundant recomputation or state storage to resolve broken causality from parallel branch validation. 
These non-monotonic state dependencies force all the intermediate hidden states to persist in either on-chip BRAM or off-chip HBM/DDR, as leaf nodes depend on them. The former incurs prohibitive on-chip memory overhead, while the latter introduces costly data transfer.

\subsection{Observation}
Sec.~\ref{subsec:mamba} identifies prohibitive storage costs from retaining all draft tree hidden states. We observe that breadth-first traversal permits state eviction once the leaf nodes' verification is complete. 
% For instance, node 1 in Figure 6a can be retained only until its children nodes 4-5 complete verification, then discarded.
For instance, $node$ $1$ in Fig.~\ref{fig:method2} (a) is retained only until its child nodes ($node$ $4$ and $node$ $5$) complete verification, after which it is discarded.
By maintaining only required active nodes, this strategy can eliminate full-tree storage.

However, storing the entire hidden state exceeds on-chip buffer capacity, which motivates tensor size reduction. Eq. \ref{eq:hidden_state_equation} reveals that $h_t$ is computed from $h_{t-1}$ via element-wise multiplication, yielding a key structural property as in Fig.~\ref{fig:method2} (b): while $h_t$ tensors exhibit inter-token data dependencies, they have no intra-token dependencies. This enables decomposition of hidden states into tiles, replacing full-tensor storage with single-tile buffering while preserving causal integrity.

\subsection{Approach}
To reduce memory overhead, we propose a hardware-friendly FIFO-based Tree Verification with Tiling. This algorithm employs breadth-first traversal and tile-wise processing. As illustrated in Fig.~\ref{fig:method2} (c), nodes are managed via an on-chip FIFO that tracks inter-token dependencies, with each node iterated $D/G$ times, where $D$ denotes hidden state dimension and $G$ denotes tile size.

Fig. \ref{fig:method2} (d) depicts execution for the tree in Fig. \ref{fig:method2} (a). At timestep $T=0$, the root node $h_0$ is loaded into the FIFO and SSM module. The SSM computes $h_1$, $h_2$, and $h_3$ at $T=1,2,3$, respectively, with $h_1$ and $h_2$ forwarded to FIFO for leaf-node processing. Non-branching nodes $h_3$ are discarded. Parent transitions trigger FIFO pops that retrieve nodes for computation in the SSM module (e.g., $T=4,6,7,9$), while cached states are reused during timesteps with unchanged parent nodes (e.g., $T=2,3,5,8$). Newly generated states are retained in the FIFO only for unresolved branches (e.g., $node$ $1,2,4,7$), while terminal nodes are discarded immediately (e.g., $node$ $3,5,6,8,9$).

The hidden states are partitioned into hardware-configurable tiles as in Fig. \ref{fig:method2} (b), with each traversal iteration processing positionally aligned tiles across all nodes. Full hidden state computation is completed in  $D/G$ iterations.
Combining node eviction with tiled execution, this approach only requires on-chip FIFO storage for at most $N/2 \times G$, where $N$ denotes node count, avoiding bulk off-chip transfers. Overlapping tiled iterations can further minimize the latency overhead.

\section{linear parallel ssm sequential dataflow}
\label{sec:method3}

% \begin{figure}[!t]

% \includegraphics[width=\linewidth]{fig/method3.pdf}

% \caption{Linear parallel SSM sequential dataflow. (a) Linear unit loads weight tiles with peak bandwidth and broadcasts them for parallel MAC operations across all tokens. (b) SSM unit unrolls multiple operators in parallel using EMUs and performs pipelined execution sequentially along the token dimension. (c) Sequential SSM execution and architecturally co-designed parallelism enable overlap with linear computation.} 
% \label{fig:method3}
% \end{figure}

\subsection{Challenge}

% Deploying the speculative decoding system on FPGA presents a critical resource utilization challenge: a naive implementation that isolates the target and draft models creates severe pipelining inefficiencies, where one model sits idle (e.g., the draft model during the target’s parallel verification phase) while the other computes, wasting over 50\% of hardware resources.

% Deploying speculative decoding systems on FPGAs faces a critical resource utilization challenge: naively isolating target and draft models introduces severe pipelining inefficiencies, where one model idles (e.g., the draft model during target verification) while the other computes, degrading hardware utilization. 

% To mitigate this, we propose a unified architecture for concurrent deployment of both models. However, this creates a workload mismatch—the compute-bound target model demands high parallelism, while the memory-bound draft model requires frequent weight accesses during low-parallelism autoregressive decoding. 

Given that the draft model and target model work alternately and share a similar computation pipeline, a unified dataflow, rather than isolating them,  is preferred to prevent hardware underutilization. However, their divergent computational demands pose challenges. The target model is compute-intensive, requiring tiled operations constrained by on-chip memory that increase SSM latency during multi-token verification. The draft model is memory-bound during sequential autoregressive token generation. Therefore, how to efficiently allocate DSP resources and schedule dataflow to simultaneously accommodate compute parallelism, memory bandwidth constraints, and overlap opportunity, remains a key design challenge.
% Conversely, a unified hardware approach that combines both models on the same FPGA improves utilization but introduces a fundamental workload mismatch—the target model demands high parallelism (e.g., 16-32 MAC units) for its compute-bound operations, while the draft model requires frequent weight access for its memory-bound, low-parallelism autoregressive decoding. 
% For improving hardware utilization, we employ a unified hardware architecture to support the deployment of both large and small models. 
% However, this approach introduces workload mismatch:
% the parallel verification of the target model and the autoregressive decoding of the draft model represent two fundamentally different workload types—the parallel verification of the target model is a compute-bound task requiring high computational parallelism, while the autoregressive decoding of the draft model is a memory-bound task with lower computational parallelism.
% the target model demands high parallelism (e.g., 32 MAC units) for its compute-bound operations, while the draft model requires frequent weight access for its memory-bound, low-parallelism (e.g., 8 MAC units) autoregressive decoding. 
% Therefore, we need to design a new dataflow to balance DSP utilization, reduce the draft model's bandwidth pressure, and overlap memory latency with computation.

\subsection{Observation}

\begin{figure}[!t]

\includegraphics[width=\linewidth]{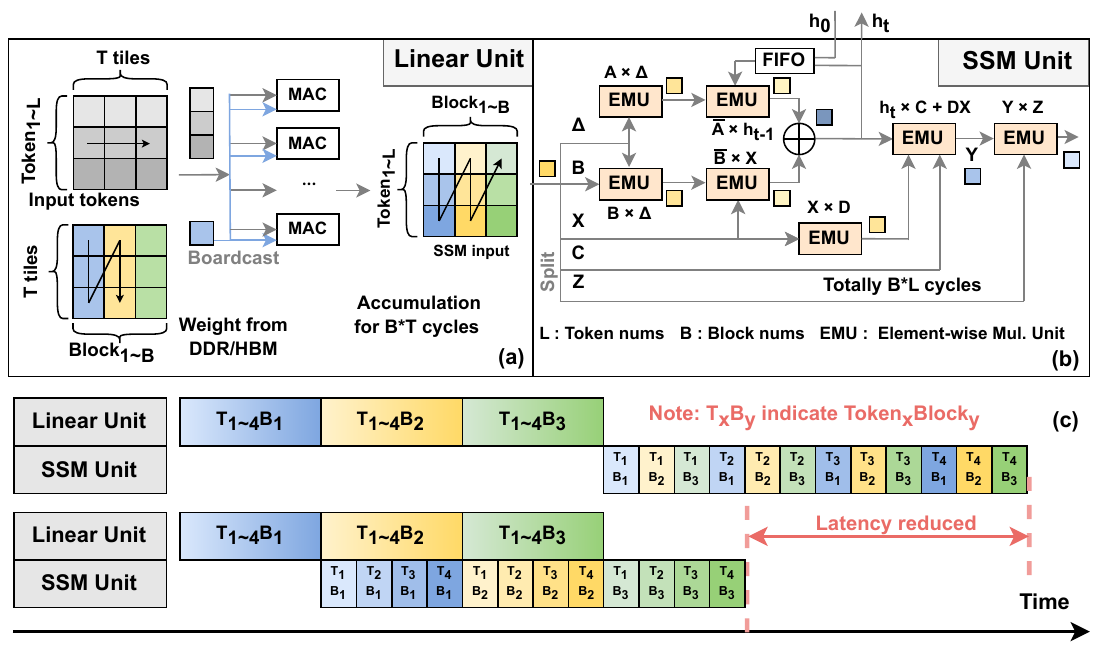}

\caption{Linear parallel SSM sequential dataflow. (a) Linear unit loads weight tiles with peak bandwidth and broadcasts them for parallel MAC operations across all tokens. (b) SSM unit unrolls multiple operators in parallel using EMUs and performs pipelined execution sequentially along the token dimension. (c) Sequential SSM execution and architecturally co-designed parallelism enable overlap with linear computation.} 
\vspace{-5pt}
\label{fig:method3}
\end{figure}

% Our key insight is to design the dataflow architecture not based on the workload patterns of target and draft models, but rather on Mamba’s operator characteristics and FPGA’s inherent bandwidth/compute constraints:

% (1) For Linear layers (MatMul), these are memory-bound, with latency dominated by off-chip weight loading. Thus, we tune parallelism based on hardware bandwidth (not workload) to avoid compute units stalling on memory access. For SSM Layers (Element-wise Ops). They are compute-bound, involving only element-wise operations on activations, with minimal off-chip traffic for $h_t$ (after optimizing via prior methods), allowing us to maximize parallelism within DSP limits.

% (2) Our latency analysis reveals a key opportunity for optimization through computational overlap. While Linear layer weights are loaded immediately, their output tiles only become available after completing multi-cycle input-channel accumulation. Crucially, this processing gap creates a natural window to interleave SSM operations, enabling full overlap between Linear computation and SSM execution. By strategically scheduling SSM computations (including both verification and autoregression phases) during these accumulation intervals, we completely hide the SSM latency behind Linear operations, achieving zero-overhead parallel processing without introducing additional pipeline stalls.

Hardware design for Mamba comprises two primary stages: linear layers and SSM layers, with the two following observations.
(1) Linear layers are memory-bound, with latency dominated by off-chip weight access. Parallelism must align with memory bandwidth to prevent compute unit stalls. SSM layers, involving compute-bound element-wise operations on activations, require minimal off-chip traffic and can maximize parallelism within DSP constraints. 
(2) Furthermore, their data dependencies differ: linear layers exhibit no inter-token dependence but require multi-tile input aggregation per output block (Fig. \ref{fig:method3} (a)), while SSM layers show inter-token dependencies with no intra-token constraints (Fig. \ref{fig:method2} (b)). This distinction enables overlap of linear and SSM computations through strategic scheduling, thus maximizing throughput.

\subsection{Approach}

Building on these insights, we design a dataflow that parallelizes linear layers across tokens while processing SSMs sequentially. This supports both parallel verification for target models and autoregressive generation for small models, as shown in Fig.~\ref{fig:method3}.

% Linear layers are computed with weight broadcasting and multi-token parallelism. Weight tiles are fetched from DDR/HBM and broadcast to processing elements (PEs), enabling concurrent multiply-and-accumulate (MAC) operations on input tiles from multiple tokens. Cross-channel accumulation per tile completes in \(T_1\) cycles, with subsequent tiles processed immediately.  

% Linear layers are computed with weight broadcasting and multi-token parallelism. Weight tiles are fetched from DDR/HBM and broadcast to processing elements (PEs), enabling concurrent multiply-and-accumulate (MAC) operations on input tiles from multiple tokens. Cross-channel accumulation per tile completes in \(T\) cycles, with subsequent tiles processed immediately.  

% SSM layers are executed in three pipelined stages with latencies: $T_2 \approx T_3 \approx T_4 < T_1$ to maximize overlap. In Stage 1, linear layer outputs are split into \(B, C, \Delta, X\) and \(Z\), processed by dedicated PEs for element-wise operations in \(T_2\) cycles. In Stage 2, \(\bar{A}\) and \(\bar{B}\) are cyclically computed with \(h_{t-1}\), generating \(h_t\) for downstream PEs and FIFOs as in (Sec.~\ref{sec:method2}).  Stage 3: \(C\) is multiplied with \(h_t\) to produce the SSM output.  

% Fig 7c illustrates our dataflow to overlap linear layer and SSM execution. Upon completing a linear layer tile, it is immediately streamed to SSM units, achieving pipeline parallelism that reduces overall latency.

For linear units in Fig. \ref{fig:method3} (a),  $L$ tokens are processed in parallel during verification. The weights are divided into $B$ blocks along the output channel, and each block is further divided into $T$ tiles along the input channel. The system maximizes its bandwidth utilization to load one tile from DDR/HBM and broadcasts it to multiple MAC units for matrix multiplication of all $L$ tokens. After $T$ cycles, the output of the first block is obtained, serving as input to the SSM Unit. The full linear unit computation is completed in $B \times T$ cycles.

As shown in Fig.~\ref{fig:method3} (b), the SSM unit comprises a variety of operators, all of which are fully unrolled and organized for pipelined execution. This unrolling is achieved through EMUs, which enable efficient parallel expansion of these operators. Once the first block of input from the Linear unit (containing $L$ tokens) becomes available, the SSM Unit can immediately begin pipelined computation. Unlike linear units, the SSM unit consumes only one tile of an individual token per cycle. Thus, it takes $L$ cycles to process one block and $L \times B$ cycles to complete the entire SSM computation.

Fig.~\ref{fig:method3} (c) demonstrates how we exploit inter-token dependencies assuming no intra-token constraints. By sequentially executing the same tile computation across different tokens in the SSM Unit and designing the parallelism such that the maximum $L$ approximates $T$, we achieve full overlap between linear and SSM computations during parallel token processing. 
% This design enables the system's overall latency to closely approach the theoretical lower bound, which is defined by the model's parameter size divided by the available memory bandwidth.
This design significantly reduces the system's overall latency.
\section{experiments}
\label{sec:experiments}

\subsection{Evaluation Setup}
% \label{subsubsec:Setup}

% \begin{table}[!t]
% \centering
% \captionsetup{justification=centering}
% \caption{Hardware Specification}
% \label{tab:hardware_para}
% \renewcommand{\arraystretch}{1.2}
% \setlength{\tabcolsep}{0.8pt}
% \begin{tabular}{
%     p{2cm}  % 第一列左对齐
%     >{\centering\arraybackslash}p{2.1cm}  % 居中
%     >{\centering\arraybackslash}p{2.1cm}  % 居中
%     >{\centering\arraybackslash}p{2.3cm}  % 居中
% }
% \toprule
%  & \makecell{\textbf{NVIDIA}\\\textbf{4090 GPU}} 
%  & \makecell{\textbf{AMD}\\\textbf{VCK190 FPGA}} 
%  & \makecell{\textbf{AMD}\\\textbf{VHK158 FPGA}} \\
% \midrule
% \textbf{Compute units} & 512 tensor cores & 1968 DSPs & 7392 DSPs \\
% \textbf{Frequency} & 2235 MHz & 400 MHz & 250 MHz \\
% \textbf{FP16 PCP} & 330 TOPS & 2.09 TOPS & 4.91 TOPS \\
% \textbf{Memory} & 24 GB & 8 GB & 32 \& 32 GB \\
% \textbf{Bandwidth} & 1008 GB/s & 12 GB/s & 819 \& 51 GB/s \\
% \bottomrule
% \end{tabular}
% \end{table}

% \begin{figure}[h]
% \centering
% \includegraphics[width=8cm,height=5.5cm]{fig/layout.pdf}
% \caption{SpecMamba implementation layout on VHK158.} 
% \label{fig:layout}
% \end{figure}

\begin{figure}[t]
\centering
\includegraphics[width=8cm,height=5.5cm]{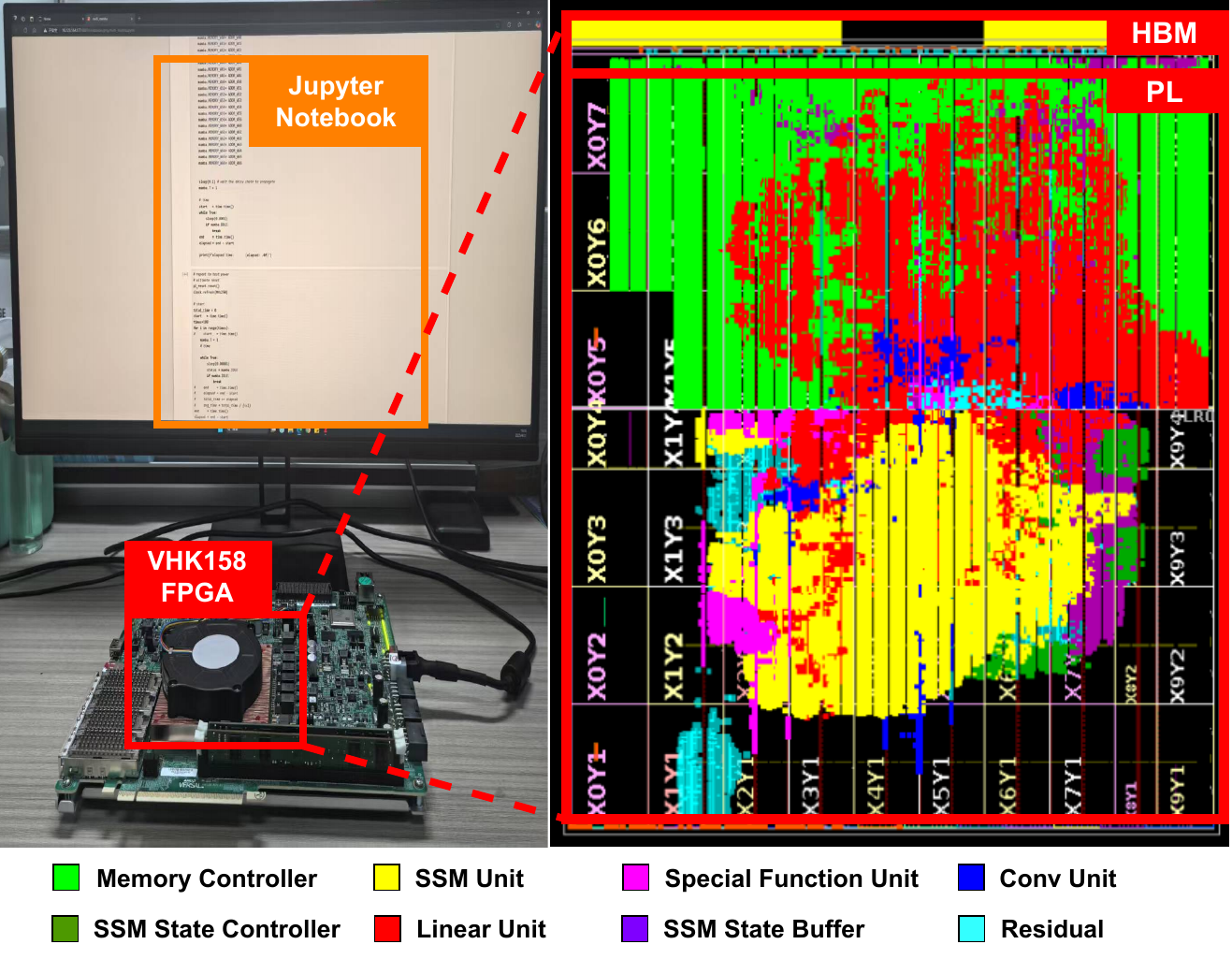}
\caption{SpecMamba implementation layout on VHK158.} 
\label{fig:layout}
\end{figure}

We evaluate the performance of SpecMamba on AMD VHK158 for HBM scenario and AMD VCK190 for DDR scenario. 
Our design is compared with several prior works, including LightMamba~\cite{lightmamba_wei2025lightmamba} on FPGA and Snakes and Ladders~\cite{SSM_Mamba_Snakes-and-Ladders_wu2024snakes} on NVIDIA RTX 4090 GPU to highlight the performance gains in terms of throughput, energy efficiency, and hardware utilization. 
The hardware specifications are detailed in Table \ref{tab:hardware_para}.
To emulate an edge device where resources are strictly constrained, we deliberately excluded the use and performance evaluation of the dedicated AI Engine available on the VCK190 platform.
SpecMamba is implemented using Vitis HLS and Vivado Design FLow on both VHK158 and VCK190. Fig.~\ref{fig:layout} shows the layout of our implementation on VHK158 FPGA, and Table~\ref{tab:resources_usage} presents its resource usage. The throughput is measured using the PYNQ framework and power consumption with the Xilinx's BEAM tool on-board.
We quantize all models to INT4 precision following \cite{lightmamba_wei2025lightmamba} and implement tree-based speculative decoding as proposed in \cite{OPT-Tree-structure_wang2025opt}.
The temperature for draft model prediction is set to 1.0, and the prediction length is set to 16 by default. Following \cite{lightmamba_wei2025lightmamba, SSM_Mamba_Snakes-and-Ladders_wu2024snakes}, we set the batch size to 1.

% \begin{figure}[h]
% \centering
% \includegraphics[width=8cm,height=5.5cm]{fig/layout.pdf}
% \caption{SpecMamba implementation layout on VHK158.} 
% \label{fig:layout}
% \end{figure}

\begin{table}[!t]
\centering
\captionsetup{justification=centering}
\caption{Hardware Specification}
\label{tab:hardware_para}
\renewcommand{\arraystretch}{1.2}
\setlength{\tabcolsep}{0.8pt}
\begin{tabular}{
    p{2cm}  % 第一列左对齐
    >{\centering\arraybackslash}p{2.1cm}  % 居中
    >{\centering\arraybackslash}p{2.1cm}  % 居中
    >{\centering\arraybackslash}p{2.3cm}  % 居中
}
\toprule
 & \makecell{\textbf{NVIDIA}\\\textbf{4090 GPU}} 
 & \makecell{\textbf{AMD}\\\textbf{VCK190 FPGA}} 
 & \makecell{\textbf{AMD}\\\textbf{VHK158 FPGA}} \\
\midrule
\textbf{Compute units} & 512 tensor cores & 1968 DSPs & 7392 DSPs \\
\textbf{Frequency} & 2235 MHz & 400 MHz & 250 MHz \\
\textbf{FP16 PCP} & 330 TOPS & 2.09 TOPS & 4.91 TOPS \\
\textbf{Memory} & 24 GB & 8 GB & 32 \& 32 GB \\
\textbf{Bandwidth} & 1008 GB/s & 12 GB/s & 819 \& 51 GB/s \\
\bottomrule
\end{tabular}
\end{table}

\textbf{Models and Datasets.} We use Mamba2-2.7B as the target model, and Mamba2-130M, 370M and 780M as draft models. We evaluate the throughput and energy efficiency on five widely adopted datasets: MT-Bench~\cite{MT-Bench_zheng2023judging}, PIQA~\cite{PIQA_bisk2020piqa}, GSM-8K~\cite{GSM8K_cobbe2021training}, Alpaca~\cite{Alpaca_taori2023stanford}, and HumanEval~\cite{HumanEval_chen2021evaluating}, which are representative of question answering, mathematical reasoning, and code generation tasks.

% \textbf{Hyperparameters.} 

% \textbf{FPGA Platforms.} 
% We use two FPGA platforms for evaluation, including the AMD VHK158 (HBM) and AMD VCK190 (DDR). Table 2~\ref{?} lists the hardware parameters of FPGA platforms. the description of table...
% \textbf{FPGA Implementation.}%
 
% \xietong{Merge with the first paragraph?}

% \textbf{GPU Baselines.} We choose NVIDIA A10G GPU (24GB HBM) and NVIDIA RTX 4090 as our GPU baselines. %, and their specifications are also listed in Table 2~\ref{?}%. 
% We use NVIDIA system management interface for power measurements. Table~\ref{tab:hardware para} shows the hardware parameters of FPGA and GPU, as well as their detailed specifications.

% \textbf{CPU Baselines.}

\begin{figure*}[!t]
\centering
\includegraphics[width=\linewidth,height=5.8cm]{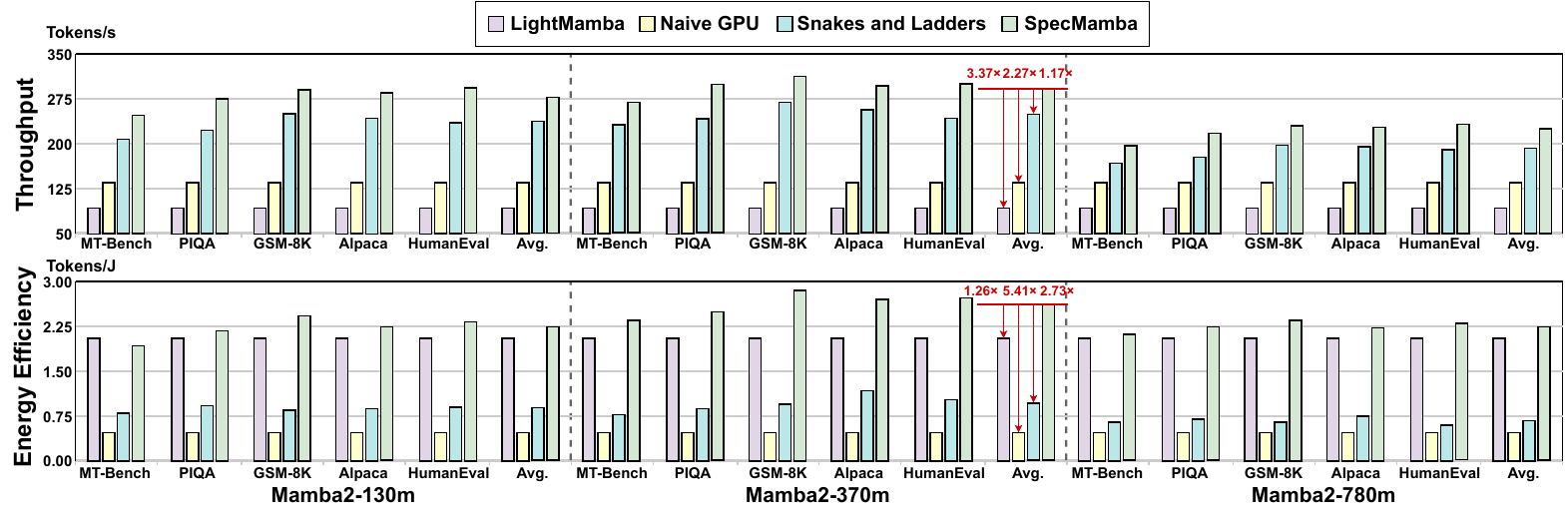}
\caption{Comparison of throughput and energy efficiency improvement of SpecMamba to LightMamba, Naive GPU, and Snakes and Ladders across three different draft model sizes.} 
\vspace{-10pt}
\label{fig:gpu}
\end{figure*}

\subsection{Evaluation Results}

\textbf{Throughput and Energy Efficiency.} Fig.~\ref{fig:gpu} presents the throughput (Tokens/s) and energy efficiency (Tokens/J) comparisons with these previous works under different draft model sizes: 
implementation on GPU without speculative decoding (Naive GPU), 
implementation on GPU with speculative decoding (Snakes) \cite{SSM_Mamba_Snakes-and-Ladders_wu2024snakes}, 
and FPGA implementation without speculative decoding (LightMamba)\cite{lightmamba_wei2025lightmamba}.
% Throughput and energy efficiency are normalized to the Naive GPU baseline. 
% Compared to Naive GPU, Naive Spec. GPU further integrates the speculative decoding strategy proposed in Snakes and Ladders~\cite{SSM_Mamba_Snakes-and-Ladders_wu2024snakes}. 
% Under a token tree length of 16, 
SpecMamba consistently outperforms both GPU baselines across all three draft model sizes: 130M, 370M, and 780M. Mamba2-370M delivers the best overall performance, 
% achieving an average improvement of xxx, 2.11×, and xxx in throughput, and xxx, 5.41×, and xxx in energy efficiency, compared to LightMamba, Naive GPU, and Naive Spec. GPU, respectively.
with average throughput improvements of 3.12$\times$, 2.11$\times$, and 1.17$\times$, and energy efficiency enhancements of 1.26$\times$, 5.41$\times$, and 2.73$\times$, over LightMamba, Naive GPU, and Snakes, respectively.
% which achieves the average xxx, 2.11× and xxx improvement in throughput and xxx, 5.41× and xxx enhancement in energy efficiency compared with the LightMamba, Naive GPU and Naive Spec. GPU, respectively. 
Compared to the 370M draft model, the 130M model delivers lower speedup, primarily due to its reduced acceptance rate, which limits the effectiveness of speculative execution. We also observe a performance drop when further increasing the model size from 370M to 780M. Although larger draft models tend to improve acceptance rates, they also introduce significantly higher computational overhead during decoding, which leads to reduced throughput and increased energy consumption.

\begin{table}[t]
    \centering
    % \captionsetup{font=small}
    \caption{Resource Usage of SpecMamba on VHK158}
    \label{tab:resources_usage}
    \renewcommand{\arraystretch}{0.9} % 控制行距
    \scriptsize                      % 缩小整体字体
    \resizebox{\linewidth}{!}{%
    \begin{tabular}{cccccc}
        \toprule
        \textbf{Component} & \textbf{LUT(k)} & \textbf{FF(k)} & \textbf{DSP} & \textbf{BRAM} & \textbf{URAM} \\
        \midrule
        \textbf{Memory Controller}       & 331.0 & 488.0 & 0    & 276 & 61  \\
        \textbf{SSM Unit}                & 253.5 & 382.1 & 1068 & 56  & 12  \\
        \textbf{SPU}                     & 35.1  & 53.6  & 193  & 163 & 12  \\
        \textbf{Conv Unit}               & 17.3  & 41.5  & 295  & 132 & 9   \\
        \textbf{SSM State Controller}    & 31.5  & 49.2  & 0    & 56  & 12  \\
        \textbf{Linear Unit}             & 254.0 & 403.4 & 3537 & 142 & 30  \\
        \textbf{SSM State Buffer}        & 17.6  & 26.9  & 0    & 415 & 231 \\
        \textbf{Residual}                & 34.9  & 53.3  & 0    & 294 & 61  \\
        \midrule
        \textbf{Total}                  & \textbf{975}   & \textbf{1498}  & \textbf{5093} & \textbf{1534}& \textbf{428} \\
        \bottomrule
    \end{tabular}%
    }
\end{table}

\begin{table}[t]
\centering
\caption{Comparison with Prior Art Works}
\label{tab:Comparison with Prior Art Works}
\renewcommand{\arraystretch}{1.2}  
\resizebox{\linewidth}{!}{%
\begin{tabular}{l|cc|cc}
\Xhline{1.2pt}
\multirow{2}{*}{\strut} & \multicolumn{2}{c|}{\textbf{HBM Scenario}} & \multicolumn{2}{c}{\textbf{DDR Scenario}} \\
                        & \textbf{LightMamba} & \textbf{Ours}     & \textbf{LightMamba} & \textbf{Ours} \\
\midrule
\textbf{Platform}       & U280 FPGA           & VHK158 FPGA       & VCK190 FPGA         & VCK190 FPGA \\
\textbf{Frequency}      & 200MHz              & 250MHz            & 400MHz              & 400MHz \\
\textbf{Bandwidth}      & 460GB/s             & 819GB/s           & 12GB/s              & 12GB/s \\
\midrule
\rowcolor{gray!20} 
\textbf{Model}          & Mamba2-2.7B         & \begin{tabular}[c]{@{}c@{}}Mamba2-2.7B\\Mamba2-370M\end{tabular}
                        & Mamba2-2.7B         & \begin{tabular}[c]{@{}c@{}}Mamba2-2.7B\\Mamba2-370M\end{tabular} \\
% \textbf{Opt.}           & INT4                  & INT4                & INT4                  & INT4 \\
\midrule
\textbf{LUT}            & 297k(22.7\%)                & 975k(56.6\%)              & 107k(11.9\%)                & 536k(59.6\%) \\
\textbf{FF}             & 394k                & 1498k             & 130k                & 643k \\
\textbf{BRAM}           & 912                 & 1534              & 912                 & 1223 \\
\textbf{DSP}            & 1164(12.8\%)                & 5093(68.8\%)              & 228(11.5\%)                 & 1106(56.2\%) \\
\textbf{URAM}           & 61                  & 428               & 61                  & 141 \\
\midrule
\rowcolor{gray!20}
\textbf{Thoughtput}     & 93                  & 313.4             & 7.21                & 20.6 \\
% \textbf{Res. Util.}     & 17.7\%                 & 62.7\%            & 11.7\%              & 57.9\% \\
\Xhline{1.2pt}
\end{tabular}%
}
\end{table}

We compare SpecMamba with the prior-art FPGA accelerator, LightMamba\cite{lightmamba_wei2025lightmamba}, across two hardware settings: HBM-FPGA and DDR-FPGA in Table~\ref{tab:Comparison with Prior Art Works}. 
Both works employ Mamba2-2.7B as the target model, while SpecMamba integrates Mamba2-370M as the draft model. For the HBM scenario, SpecMamba achieves a practical throughput of 313.4 tokens/s on the VHK158 FPGA, which outperforms LightMamba on the U280 FPGA by 3.37$\times$ on average.
% representing a 3.37× and 2.46× speedup over LightMamba on the U280 FPGA. 
% Compared to LightMamba, SpecMamba achieves higher on-chip computing resource utilization, reaching 62.7\%. 
For the DDR scenario, SpecMamba achieves a throughput of 20.6 tokens/s 
% and a compute resource utilization of 57.9\%
on the VCK190 FPGA, which is 2.85$\times$ 
% and 3.37$\times$ 
higher than 
% those of 
LightMamba. 
% respectively. 
SpecMamba consistently outperforms prior works in both high-bandwidth and bandwidth-limited conditions.

% \begin{table}[!t]
% \centering
%     \caption{Average number of tokens accepted in a decoding step. We use Mamba2-2.7B and Mamba2-370M as the target and draft model.}
%     \renewcommand{\arraystretch}{1.2}  
%     \label{tab:accept length}
% \begin{tabular}{cc|cccccc}
% \Xhline{0.9pt}
%        \multirow{2}{*}{\makecell{\textbf{Draft}\\\textbf{Structure}}}           &  \multirow{2}{*}{\textbf{Dataset}}          & \multicolumn{6}{c}{\textbf{Prediction Lengths}}    \\
%                           &    & 6    & 8    & 10    & 12    & 14   & 16  \\ \hline
% \multirow{3}{*}{\textbf{Sequence}} & MT-Bench  & 1.78 & 1.97 & 2.24 & 2.39 & 2.58 & 2.71 \\
%                           & GSM-8K    & 2.12 & 2.33 & 2.86 & 2.95 & 3.14 & 3.17 \\
%                           & HumanEval & 2.05 & 2.28 & 2.68 & 2.89 & 3.04 & 3.09 \\ \hline
% \multirow{3}{*}{\textbf{Tree}}     & MT-Bench  & 3.11 & 3.63 & 3.92 & 4.16 & 4.77 & 4.91 \\
%                           & GSM-8K    & 3.82 & 4.41 & 4.81 & 5.03 & 5.94 & 5.98 \\
%                           & HumanEval & 3.69 & 4.27 & 4.69 & 4.92 & 5.76 & 5.82 \\ 
% \Xhline{0.9pt}                         
% \end{tabular}
% \end{table}

\begin{table}[htbp]
\centering
\caption{Average tokens accepted per decoding step using Mamba2-2.7B as target and Mamba2-370M as draft model.}
\renewcommand{\arraystretch}{1.2}  
\label{tab:accept length}
\begin{tabular}{cc|cccccc}
\Xhline{0.9pt}
       \multirow{2}{*}{\makecell{\textbf{Draft}\\\textbf{Structure}}}           &  \multirow{2}{*}{\textbf{Dataset}}          & \multicolumn{6}{c}{\textbf{Prediction Lengths}}    \\
                          &    & 6    & 8    & 10    & 12    & 14   & 16  \\ \hline
\multirow{3}{*}{\textbf{Sequence}} & MT-Bench  & 1.78 & 1.97 & 2.24 & 2.39 & 2.58 & 2.71 \\
                          & GSM-8K    & 2.12 & 2.33 & 2.86 & 2.95 & 3.14 & 3.17 \\
                          & HumanEval & 2.05 & 2.28 & 2.68 & 2.89 & 3.04 & 3.09 \\ \hline
\multirow{3}{*}{\textbf{Tree}}     & MT-Bench  & 3.11 & 3.63 & 3.92 & 4.16 & 4.77 & 4.91 \\
                          & GSM-8K    & 3.82 & 4.41 & 4.81 & 5.03 & 5.94 & 5.98 \\
                          & HumanEval & 3.69 & 4.27 & 4.69 & 4.92 & 5.76 & 5.82 \\ 
\Xhline{0.9pt}                         
\end{tabular}
\end{table}

\textbf{Acceptance Rate.} In Table~\ref{tab:accept length},
we also provide the acceptance rates under different prediction lengths. Across all settings, our tree-based speculative decoding consistently achieves a higher acceptance rate compared to the sequence-based approach across all prediction lengths. Moreover, we also observe that the acceptance rate increases with prediction lengths, indicating more efficient draft token utilization in longer sequences.

% \textbf{T1 results}

% \begin{table}[h]
%     \centering
%     \caption{Comparison of Different Methods}
%     \renewcommand{\arraystretch}{1.0}  
%     \label{tab:T1 Results Comparison of Different Methods}
%     \begin{tabularx}{\columnwidth}{l >{\centering\arraybackslash}X >{\centering\arraybackslash}X >{\centering\arraybackslash}X >{\centering\arraybackslash}X}
%     \toprule
%      & \makecell{\bfseries Compute \\ \bfseries Growth } 
%        & \makecell{\bfseries DSP } 
%        & \makecell{\bfseries Memory \\ \bfseries Access } 
%        & \makecell{\bfseries Latency \\ \bfseries Increase} \\
%     \midrule
%     \textbf{Method 1} & $\times 1$      & $\times 1$      & $\times 2$      & $\times 2$    \\
%     \textbf{Method 2} & $\times 2\sim4$  & $\times 4$  & $\times 1$      & $\times 1$    \\
%     \textbf{Method 3} & $\times 1\sim2$  & $\times 2$      & $\times 1$  & $\times 1$    \\
%     \textbf{Method 4} & $\times 1.1$     & $\times 1.1$     & $\times 1.1$      & $\times 1$  \\
%     \bottomrule
%     \end{tabularx}
% \end{table}

% \textbf{T2 results}
% \textbf{T3 results}

\begin{figure}[!t]
\centering
\includegraphics[width=\linewidth]{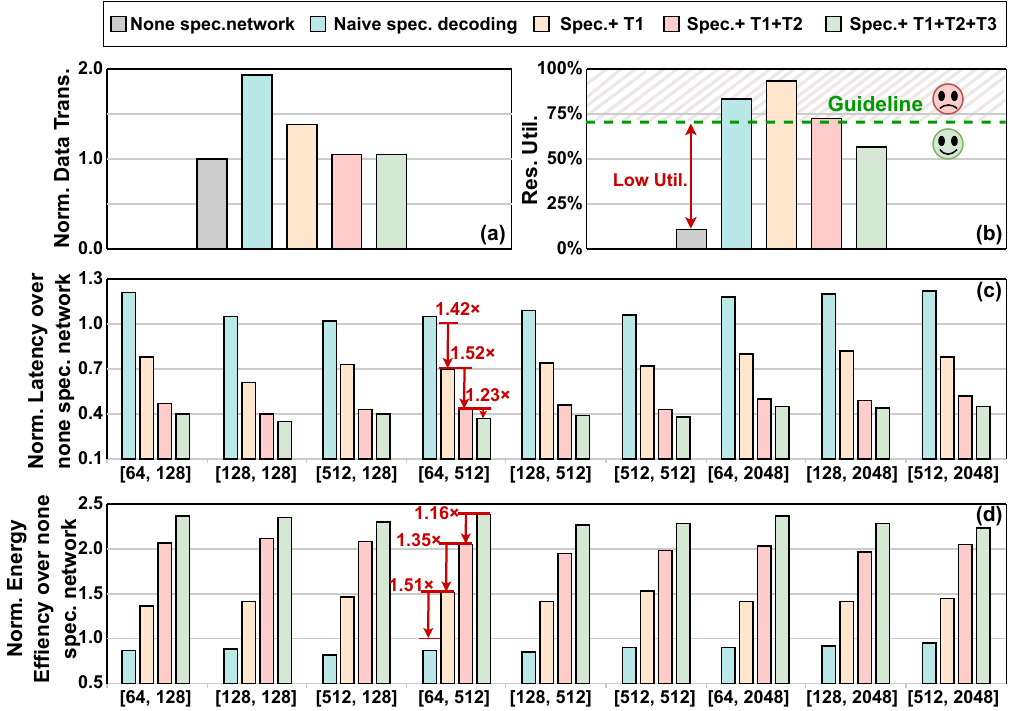}
\caption{Ablation study: (a) Normalized data transmission. (b) Resource utilization. (c) Normalized latency over None spec. network. (d) Normalized energy efficiency over None spec. network. T1: Memory-aware hybrid backtracking, T2: FIFO-based tree verification with tiling, T3: Linear parallel SSM sequential dataflow.} 
\label{fig:ablation}
\vspace{-10pt} 
\end{figure}

\textbf{Ablation Study.} We conduct an ablation study on SpecMamba on the VCK190 platform to evaluate the impact of three different techniques on data transmission, resource consumption, latency, and energy efficiency in Fig.~\ref{fig:ablation}.

% In Fig.~\ref{fig:ablation}, "None Spec. Network" refers to the baseline that directly deploys the full Mamba2-2.7B model on FPGA without speculative decoding. "Naive Spec. Decoding" denotes a speculative decoding setup without optimization. Our method, SpecMamba, builds on the naive baseline with three key techniques: 

Fig.~\ref{fig:ablation} (a) shows the effect of our methods on data transmission reduction. The Naive Spec. Decoding results in the highest transmission cost, as it stores all intermediate hidden states off-chip, leading to significant data movement. In contrast, our proposed techniques progressively reduce the transmission volume, bringing it closer to that of the normalized None Spec. Network baseline.

% \begin{figure}[!t]
% \centering
% \includegraphics[width=0.95\linewidth]{fig/ablation.pdf}
% \caption{Ablation study: (a) Normalized data transmission. (b) Resource utilization. (c) Normalized latency over None spec. network. (d) Normalized energy efficiency over None spec. network. T1: Memory-aware hybrid backtracking, T2: FIFO-based tree verification with tiling, T3: Linear parallel SSM sequential dataflow.} 
% \label{fig:ablation}
% \vspace{-10pt} 
% \end{figure}

In Fig.~\ref{fig:ablation} (b), we illustrate the chip resource utilization of different methods, defined as the average utilization rate of on-chip LUT and DSP resources. 
% the utilization of computing resources on the chip of different methods.
% More specifically, it is defined as the average utilization rate of on-chip LUT and DSP resources.
Note that a widely used guideline for FPGA resource utilization is 70\%, above which may potentially affect timing and routing.
The None spec. network suffers from bandwidth limitations, resulting in low utilization far below the 70\% guideline. Naive spec. decoding and Spec. +T1 causes overly high utilization, which could potentially impact timing and routing. Incorporation of T2 and T3 reduces utilization to 57.9\%, keeping it below the guideline and making the design more routing-friendly.

Fig.~\ref{fig:ablation} (c) and (d) show the normalized latency and energy efficiency over the None spec. network using our techniques, across various prefill and decoding lengths. For example, for a prefill length of 64 and a decode stage of 512, T1, T2, T3 reduce the latency by 1.42$\times$, 1.52$\times$, 1.23$\times$, and improve the energy efficiency by 1.51$\times$, 1.35$\times$, 1.16$\times$, respectively, demonstrating a gradual reduction in latency.

\section{conclusion}
\label{sec:conclusion}

This work presents SpecMamba, the first FPGA-accelerated framework enabling efficient Mamba inference with speculative decoding for edge deployment. 
% In order to balance the sequential processing of SSM and the parallel verification of speculative decoding,
% we propose a cross-layer optimization framework: 
We propose a cross-layer optimization framework to balance the sequential processing of SSMs and parallel verification of speculative decoding: 
(1) a memory-aware hybrid backtracking strategy that coordinates both models 
to resolve hidden state recovery bottlenecks, (2) a FIFO-based tree verification scheme 
% based on FIFO 
with a tiling algorithm that enables full candidate token verification in SSMs, 
% through the exploitation of intertoken dependency, 
and (3) a linear parallel and SSM sequential dataflow that achieves balancing of the model workload through hardware-software co-design.
% Implemented on AMD VHK158 and VCK190 FPGAs, SpecMamba demonstrates significant improvements over existing solutions. It achieves 2.26$\times$ speedup over GPU baselines and 2.85$\times$ improvement compared to prior FPGA implementations. 
% Furthermore, our solution shows 5.41$\times$ and 1.28$\times$ higher energy efficiency, respectively. 
Implemented on AMD VHK158 and VCK190 FPGAs, SpecMamba achieves 2.27× speedup over GPU baselines and 2.85× over prior FPGA designs, with 5.41× and 1.26× higher energy efficiency, respectively.
These results validate SSMs for high-throughput edge applications and establish a new co-design paradigm for sequential state-space architectures.

\FloatBarrier             % 确保图表不影响参考文献
\clearpage
\bibliographystyle{IEEEtran}
\bibliography{reference}

% Generated by IEEEtran.bst, version: 1.14 (2015/08/26)
\begin{thebibliography}{10}
\providecommand{\url}[1]{#1}
\csname url@samestyle\endcsname
\providecommand{\newblock}{\relax}
\providecommand{\bibinfo}[2]{#2}
\providecommand{\BIBentrySTDinterwordspacing}{\spaceskip=0pt\relax}
\providecommand{\BIBentryALTinterwordstretchfactor}{4}
\providecommand{\BIBentryALTinterwordspacing}{\spaceskip=\fontdimen2\font plus
\BIBentryALTinterwordstretchfactor\fontdimen3\font minus \fontdimen4\font\relax}
\providecommand{\BIBforeignlanguage}[2]{{%
\expandafter\ifx\csname l@#1\endcsname\relax
\typeout{** WARNING: IEEEtran.bst: No hyphenation pattern has been}%
\typeout{** loaded for the language `#1'. Using the pattern for}%
\typeout{** the default language instead.}%
\else
\language=\csname l@#1\endcsname
\fi
#2}}
\providecommand{\BIBdecl}{\relax}
\BIBdecl

\bibitem{Mamba1_gu2023mamba}
A.~Gu and T.~Dao, ``Mamba: Linear-time sequence modeling with selective state spaces,'' \emph{arXiv preprint arXiv:2312.00752}, 2023.

\bibitem{Mamba2_dao2024transformersareSSMs}
T.~Dao and A.~Gu, ``Transformers are ssms: Generalized models and efficient algorithms through structured state space duality,'' \emph{arXiv preprint arXiv:2405.21060}, 2024.

\bibitem{touvron2023llamaopenefficientfoundation}
\BIBentryALTinterwordspacing
H.~Touvron, T.~Lavril, G.~Izacard, X.~Martinet, M.-A. Lachaux, T.~Lacroix, B.~Rozière, N.~Goyal, E.~Hambro, F.~Azhar, A.~Rodriguez, A.~Joulin, E.~Grave, and G.~Lample, ``Llama: Open and efficient foundation language models,'' 2023. [Online]. Available: \url{https://arxiv.org/abs/2302.13971}
\BIBentrySTDinterwordspacing

\bibitem{OPT_zhang2022opt}
S.~Zhang, S.~Roller, N.~Goyal, M.~Artetxe, M.~Chen, S.~Chen, C.~Dewan, M.~Diab, X.~Li, X.~V. Lin \emph{et~al.}, ``Opt: Open pre-trained transformer language models,'' \emph{arXiv preprint arXiv:2205.01068}, 2022.

\bibitem{Vision_liu2024vmamba}
Y.~Liu, Y.~Tian, Y.~Zhao, H.~Yu, L.~Xie, Y.~Wang, Q.~Ye, J.~Jiao, and Y.~Liu, ``Vmamba: Visual state space model,'' \emph{Advances in neural information processing systems}, vol.~37, pp. 103\,031--103\,063, 2024.

\bibitem{Vision_yang2024plainmamba}
C.~Yang, Z.~Chen, M.~Espinosa, L.~Ericsson, Z.~Wang, J.~Liu, and E.~J. Crowley, ``Plainmamba: Improving non-hierarchical mamba in visual recognition,'' \emph{arXiv preprint arXiv:2403.17695}, 2024.

\bibitem{Vision_zhu2024visionmambaefficientvisual}
\BIBentryALTinterwordspacing
L.~Zhu, B.~Liao, Q.~Zhang, X.~Wang, W.~Liu, and X.~Wang, ``Vision mamba: Efficient visual representation learning with bidirectional state space model,'' 2024. [Online]. Available: \url{https://arxiv.org/abs/2401.09417}
\BIBentrySTDinterwordspacing

\bibitem{Multi-Modal_wan2025sigma}
Z.~Wan, P.~Zhang, Y.~Wang, S.~Yong, S.~Stepputtis, K.~Sycara, and Y.~Xie, ``Sigma: Siamese mamba network for multi-modal semantic segmentation,'' in \emph{2025 IEEE/CVF Winter Conference on Applications of Computer Vision (WACV)}.\hskip 1em plus 0.5em minus 0.4em\relax IEEE, 2025, pp. 1734--1744.

\bibitem{Multi-Modal_zhao2024cobra}
H.~Zhao, M.~Zhang, W.~Zhao, P.~Ding, S.~Huang, and D.~Wang, ``Cobra: Extending mamba to multi-modal large language model for efficient inference,'' \emph{arXiv preprint arXiv:2403.14520}, 2024.

\bibitem{Multi-Modal_dong2024fusionmamba}
W.~Dong, H.~Zhu, S.~Lin, X.~Luo, Y.~Shen, X.~Liu, J.~Zhang, G.~Guo, and B.~Zhang, ``Fusion-mamba for cross-modality object detection,'' \emph{arXiv preprint arXiv:2404.09146}, 2024.

\bibitem{Medicalwang2024mamba}
Z.~Wang, J.-Q. Zheng, Y.~Zhang, G.~Cui, and L.~Li, ``Mamba-unet: Unet-like pure visual mamba for medical image segmentation,'' \emph{arXiv preprint arXiv:2402.05079}, 2024.

\bibitem{Medicalyang2024vivim}
Y.~Yang, Z.~Xing, L.~Yu, C.~Huang, H.~Fu, and L.~Zhu, ``Vivim: A video vision mamba for medical video segmentation,'' \emph{arXiv preprint arXiv:2401.14168}, 2024.

\bibitem{Medicalguo2024mambamorph}
T.~Guo, Y.~Wang, S.~Shu, D.~Chen, Z.~Tang, C.~Meng, and X.~Bai, ``Mambamorph: a mamba-based framework for medical mr-ct deformable registration,'' \emph{arXiv preprint arXiv:2401.13934}, 2024.

\bibitem{Others3D_medical_image_xing2024segmamba}
Z.~Xing, T.~Ye, Y.~Yang, G.~Liu, and L.~Zhu, ``Segmamba: Long-range sequential modeling mamba for 3d medical image segmentation,'' in \emph{International Conference on Medical Image Computing and Computer-Assisted Intervention}.\hskip 1em plus 0.5em minus 0.4em\relax Springer, 2024, pp. 578--588.

\bibitem{PointCloudli20243dmambacomplete}
Y.~Li, W.~Yang, and B.~Fei, ``3dmambacomplete: Exploring structured state space model for point cloud completion,'' \emph{arXiv preprint arXiv:2404.07106}, 2024.

\bibitem{PointCloud_zhang2024point}
T.~Zhang, H.~Yuan, L.~Qi, J.~Zhang, Q.~Zhou, S.~Ji, S.~Yan, and X.~Li, ``Point cloud mamba: Point cloud learning via state space model,'' \emph{arXiv preprint arXiv:2403.00762}, 2024.

\bibitem{PointCloud_liang2024pointmamba}
D.~Liang, X.~Zhou, W.~Xu, X.~Zhu, Z.~Zou, X.~Ye, X.~Tan, and X.~Bai, ``Pointmamba: A simple state space model for point cloud analysis,'' \emph{arXiv preprint arXiv:2402.10739}, 2024.

\bibitem{Speculative_leviathan2023Fast-Inference-via-speculative-decoding}
Y.~Leviathan, M.~Kalman, and Y.~Matias, ``Fast inference from transformers via speculative decoding,'' in \emph{International Conference on Machine Learning}.\hskip 1em plus 0.5em minus 0.4em\relax PMLR, 2023, pp. 19\,274--19\,286.

\bibitem{Speculative_chen2023accelerating-large-language-model-decoding-with-speculative-sampling}
C.~Chen, S.~Borgeaud, G.~Irving, J.-B. Lespiau, L.~Sifre, and J.~Jumper, ``Accelerating large language model decoding with speculative sampling,'' \emph{arXiv preprint arXiv:2302.01318}, 2023.

\bibitem{Speculative_miao2024_specinfer}
X.~Miao, G.~Oliaro, Z.~Zhang, X.~Cheng, Z.~Wang, Z.~Zhang, R.~Y.~Y. Wong, A.~Zhu, L.~Yang, X.~Shi \emph{et~al.}, ``Specinfer: Accelerating large language model serving with tree-based speculative inference and verification,'' in \emph{Proceedings of the 29th ACM International Conference on Architectural Support for Programming Languages and Operating Systems, Volume 3}, 2024, pp. 932--949.

\bibitem{SSM_Mamba_Snakes-and-Ladders_wu2024snakes}
Y.~Wu, Y.~Dukler, M.~Trager, A.~Achille, W.~Xia, and S.~Soatto, ``Snakes and ladders: Accelerating state space model inference with speculative decoding,'' 2024.

\bibitem{Mamba-in-llamba_wang2024mamba}
J.~Wang, D.~Paliotta, A.~May, A.~Rush, and T.~Dao, ``The mamba in the llama: Distilling and accelerating hybrid models,'' \emph{Advances in Neural Information Processing Systems}, vol.~37, pp. 62\,432--62\,457, 2024.

\bibitem{Mamba_MARCA_li2024marca}
J.~Li, S.~Huang, J.~Xu, J.~Liu, L.~Ding, N.~Xu, and G.~Dai, ``Marca: Mamba accelerator with reconfigurable architecture,'' in \emph{Proceedings of the 43rd IEEE/ACM International Conference on Computer-Aided Design}, 2024, pp. 1--9.

\bibitem{lightmamba_wei2025lightmamba}
R.~Wei, S.~Xu, L.~Zhong, Z.~Yang, Q.~Guo, Y.~Wang, R.~Wang, and M.~Li, ``Lightmamba: Efficient mamba acceleration on fpga with quantization and hardware co-design,'' \emph{arXiv preprint arXiv:2502.15260}, 2025.

\bibitem{OPT-Tree-structure_wang2025opt}
J.~Wang, Y.~Su, J.~Li, Q.~Xia, Z.~Ye, X.~Duan, Z.~Wang, and M.~Zhang, ``Opt-tree: Speculative decoding with adaptive draft tree structure,'' \emph{Transactions of the Association for Computational Linguistics}, vol.~13, pp. 188--199, 2025.

\bibitem{li2024eagle1}
Y.~Li, F.~Wei, C.~Zhang, and H.~Zhang, ``Eagle: Speculative sampling requires rethinking feature uncertainty,'' \emph{arXiv preprint arXiv:2401.15077}, 2024.

\bibitem{li2024eagle2}
------, ``Eagle-2: Faster inference of language models with dynamic draft trees,'' \emph{arXiv preprint arXiv:2406.16858}, 2024.

\bibitem{li2025eagle3}
------, ``Eagle-3: Scaling up inference acceleration of large language models via training-time test,'' \emph{arXiv preprint arXiv:2503.01840}, 2025.

\bibitem{cai2024medusa}
T.~Cai, Y.~Li, Z.~Geng, H.~Peng, J.~D. Lee, D.~Chen, and T.~Dao, ``Medusa: Simple llm inference acceleration framework with multiple decoding heads,'' \emph{arXiv preprint arXiv:2401.10774}, 2024.

\bibitem{FPGA-accelerator_zeng2024flightllm_daiguohao}
S.~Zeng, J.~Liu, G.~Dai, X.~Yang, T.~Fu, H.~Wang, W.~Ma, H.~Sun, S.~Li, Z.~Huang \emph{et~al.}, ``Flightllm: Efficient large language model inference with a complete mapping flow on fpgas,'' in \emph{Proceedings of the 2024 ACM/SIGDA International Symposium on Field Programmable Gate Arrays}, 2024, pp. 223--234.

\bibitem{background_Pushingup_li2025pushing}
J.~Li, T.~Li, G.~Shen, D.~Zhao, Q.~Zhang, and Y.~Zeng, ``Pushing up to the limit of memory bandwidth and capacity utilization for efficient llm decoding on embedded fpga,'' \emph{arXiv preprint arXiv:2502.10659}, 2025.

\bibitem{Edgellm_huang2025edgellm}
M.~Huang, A.~Shen, K.~Li, H.~Peng, B.~Li, Y.~Su, and H.~Yu, ``Edgellm: A highly efficient cpu-fpga heterogeneous edge accelerator for large language models,'' \emph{IEEE Transactions on Circuits and Systems I: Regular Papers}, 2025.

\bibitem{MEADOW_moitra2025meadow}
A.~Moitra, A.~Ghosh, S.~Agarwal, A.~Amarnath, K.~Swaminathan, and P.~Panda, ``Meadow: Memory-efficient dataflow and data packing for low power edge llms,'' \emph{arXiv preprint arXiv:2503.11663}, 2025.

\bibitem{MT-Bench_zheng2023judging}
L.~Zheng, W.-L. Chiang, Y.~Sheng, S.~Zhuang, Z.~Wu, Y.~Zhuang, Z.~Lin, Z.~Li, D.~Li, E.~Xing \emph{et~al.}, ``Judging llm-as-a-judge with mt-bench and chatbot arena,'' \emph{Advances in Neural Information Processing Systems}, vol.~36, pp. 46\,595--46\,623, 2023.

\bibitem{PIQA_bisk2020piqa}
Y.~Bisk, R.~Zellers, J.~Gao, Y.~Choi \emph{et~al.}, ``Piqa: Reasoning about physical commonsense in natural language,'' in \emph{Proceedings of the AAAI conference on artificial intelligence}, vol.~34, no.~05, 2020, pp. 7432--7439.

\bibitem{GSM8K_cobbe2021training}
K.~Cobbe, V.~Kosaraju, M.~Bavarian, M.~Chen, H.~Jun, L.~Kaiser, M.~Plappert, J.~Tworek, J.~Hilton, R.~Nakano \emph{et~al.}, ``Training verifiers to solve math word problems, 2021,'' \emph{URL https://arxiv. org/abs/2110.14168}, vol.~9, 2021.

\bibitem{Alpaca_taori2023stanford}
R.~Taori, I.~Gulrajani, T.~Zhang, Y.~Dubois, X.~Li, C.~Guestrin, P.~Liang, and T.~B. Hashimoto, ``Stanford alpaca: An instruction-following llama model,'' 2023.

\bibitem{HumanEval_chen2021evaluating}
M.~Chen, J.~Tworek, H.~Jun, Q.~Yuan, H.~P. D.~O. Pinto, J.~Kaplan, H.~Edwards, Y.~Burda, N.~Joseph, G.~Brockman \emph{et~al.}, ``Evaluating large language models trained on code,'' \emph{arXiv preprint arXiv:2107.03374}, 2021.

\end{thebibliography}

\end{document}